\documentclass[prb,twocolumn,showpacs,superscriptaddress]{revtex4}

\usepackage{amsmath}
\usepackage{amssymb}
\usepackage{graphicx}
\usepackage[hypertex]{hyperref}
\usepackage{graphicx}
\usepackage{srctex}
\usepackage[usenames]{color}
\usepackage{dcolumn}
\usepackage{bm}

\definecolor{MidnightBlue}{cmyk}{0.98,0.13,0,0.43}
\definecolor{DarkGreen}{rgb}{0,0.7,0.1}

\begin{document}
\newcommand{\remark}[1] {\noindent\framebox{
\begin{minipage}{0.96\columnwidth}\textbf{\textit{ #1}}
\end{minipage}}
}

\newcommand{\bnabla}{{\boldsymbol{\nabla}}} \newcommand{\Tr}{\mathrm{Tr}} \newcommand{\Dk}{\check{\Delta}_{\cal K}} \newcommand{\Qk}{\check{Q}_{\cal K}} \newcommand{\Fk}{\check{\Phi}_{\cal K}}
\newcommand{\fk}{\check{\phi}_{\cal K}} \newcommand{\Ak}{\check{\mathbf{A}}_{\cal K}} \newcommand{\Si}{\check{\Xi}}
\newcommand{\cK}{\cal K}
\newcommand{\Lk}{\check{\Lambda}}
\newcommand{\bz}{{\mathbf z}}
\newcommand{\bx}{{\mathbf x}}
\newcommand{\br}{{\mathbf r}}
\newcommand{\bG}{{\mathbf G}}
\newcommand{\bu}{{\mathbf u}}
\newcommand{\bq}{{\mathbf q}}
\newcommand{\cH}{{\cal H}}
\newcommand{\dif}{{\mathrm d}}

\def\Xint#1{\mathchoice
{\XXint\displaystyle\textstyle{#1}}%
{\XXint\textstyle\scriptstyle{#1}}%
{\XXint\scriptstyle\scriptscriptstyle{#1}}%
{\XXint\scriptscriptstyle\scriptscriptstyle{#1}}%
\!\int}
\def\XXint#1#2#3{{\setbox0=\hbox{$#1{#2#3}{\int}$}
\vcenter{\hbox{$#2#3$}}\kern-.5\wd0}}
\def\ddashint{\Xint=}
\def\dashint{\Xint-}

\title{Theoretical proposal for dual transformation between the Josephson effect and 
quantum phase slip in single junction systems and nanowires}
\author{M.~Yoneda}
\affiliation{Aichi University of Technology, 50-2 Umanori Nishihasama-cho, Aichi 443-0047, Japan}
\author{M.~Niwa}
\author{N.~Hirata}
\author{M.~Motohashi}
\affiliation{School of Engineering, Tokyo Denki University, 5 Senju Asahi-cho, Adachi-ku Tokyo 120-855, Japan}

\date{\today}

\begin{abstract}
A method was devised to construct a generalized dual field theory in the quantum field theory. As a simple example using this method, we examined the duality between coherent quantum phase slip and the Josephson effect in single junction systems and nanowires. The this method was proved to be reliable within the Villain approximation.
\end{abstract}



\maketitle
\section *{\large{Introduction}}
The dual transformation has been known to be a useful tool in various physical systems. In particular, in quantum field theory \cite{ref1}-\cite{ref5} and statistical mechanics \cite{ref4}-\cite{ref7}, many studied cases incorporating duality are known.
Similarly, electric circuits arranged in series and parallel within classical electrical engineering exhibit similar adding laws, which are satisfied by interchanging the role of resistance and conductance, inductance and capacitance, current and voltage. This rule is known as the duality principle of electrical circuits\cite{ref8}-\cite{ref10}. In recent years, numerous experiments and theoretical discussions have been conducted on the potential of quantum phase-slip ($Q\!P\!S$) existing as a dual system in the Josephson junction ($J\!J$) system using nanowires \cite{ref11}-\cite{ref18}. However, a deterministic experimental fact showing the existence of quantum phase-slip completely has not been found yet. Also, the exact theory of the dual transformation between the $J\!J$ system and the $Q\!P\!S$ junction ($Q\!P\!S\!J$) system has not been completed yet \cite{ref16}-\cite{ref18}. 
 In this paper, we introduce two Hamiltonian, which are dual to each other, and propose a general theory to construct a dual system by applying the dual condition between current and voltage in an electric circuit. This method was named the dual Hamiltonian ($D\!H$) method. By using this method, the Hamiltonians of the $Q\!P\!S$ system and the $J\!J$ system are proved to be equivalent to each other by dual transformation and also prove to be an exact dual system. The remainder of this paper is organized as follows. In the next section, the $D\!H$ method is applied to build a quantum $L\!C$ circuit as a simple example. In section 2, as an application of the preceding section, the relationship between the $Q\!P\!S$ system and the $J\!J$ system in a single junction is introduced. In section 3, self-duality in various quantum junction circuits is briefly proved. In the section 4, superconductors and superinsulators are discussed from the standpoint of quantum phase transition. In section 5, duality is examined for the partition function of a single junction that incorporates quantum effects using path integration. In section 6, the derivation of the anisotropic \scalebox{0.9}{$XY(A\!X\!Y)$} model and dual anisotropic \scalebox{0.9}{$XY(D\!A\!X\!Y)$} model are described in the classical $1\! + \!1$ dimensional system equivalent to the $J\!J$ and $Q\!P\!S\!J$ in a nanowire, which is a quantum one-dimensional system. In section 7, the duality between the \scalebox{0.9}{$A\!X\!Y$} model and \scalebox{0.9}{$D\!A\!X\!Y$} model is proved by the Villain approximation. In  section  8,  we derived Ginzburg-Landau theories of two types and compared each of their critical values with the critical values in the Kosterlitz-Thouless theory. In the last section, the summary, discussion, and conclusions are presented.
\setcounter{section}{0}

\section{DH method in the quantum LC circuit \label{sec1}}

In this section, examples of quantum $L\!C$ circuits \cite{ref19} are presented as the simplest application of the $D\!H$ method. First, the Hamiltonian of a normal quantum $L\!C$ circuit is introduced as follows:
\begin{align}\label{eq1}
\scalebox{0.85}{$\displaystyle
{{H}_{LC}}=\frac{{{Q}^{2}}\left( t \right)}{2C}+\frac{{{\Phi }^{2}}\left( t \right)}{2L}
$},
\end{align}
where \scalebox{0.9}{$C$} and \scalebox{0.9}{$L$} are the capacitance and inductance of a quantum $L\!C$ circuit, respectively. The commutation relationship between the electric charge \scalebox{0.9}{$\displaystyle Q\left(x\right)$} and magnetic flux \scalebox{0.9}{$\displaystyle \Phi\left(x\right)$} is described as follows: 
\begin{align}\label{eq2}
\scalebox{0.85}{$\displaystyle
\Bigl[ \Phi \left( t \right),Q\left( t \right) \Bigr]=i\hbar
$}.
\end{align}
From the Hamiltonian of Eq.(\ref{eq1}), the equations of motion are given by:
\begin{align}\label{eq3}
\scalebox{0.85}{$\displaystyle
I\left( t \right)=\frac{dQ\left( t \right)}{dt}=\frac{i}{\hbar }\Bigl[ {{H}_{LC}},Q\left( t \right) \Bigr]=\frac{-\Phi \left( t \right)}{L}
$},\nonumber \\
\scalebox{0.85}{$\displaystyle
V\left( t \right)=-\frac{d\Phi \left( t \right)}{dt}=-\frac{i}{\hbar }\Bigl[ {{H}_{LC}},\Phi \left( t \right) \Bigr]=-\frac{Q\left( t \right)}{C}
$},
\end{align}
where \scalebox{0.9}{$I\left(t\right)$} and \scalebox{0.9}{$V\left(t\right)$} are the current and voltage of an $L\!C$ circuit, respectively. The dual Hamiltonian \scalebox{0.9}{${{\tilde{H}}_{LC}}$} is then introduced for the quantum $L\!C$ circuit, assuming the following:
\begin{align}\label{eq4}
\scalebox{0.85}{$\displaystyle
{{\tilde{H}}_{LC}}=\frac{{{{\tilde{Q}}}^{2}}\left( t \right)}{2\tilde{C}}+\frac{{{{\tilde{\Phi }}}^{2}}\left( t \right)}{2\tilde{L}}
$},
\end{align}
where \scalebox{0.9}{${\tilde{C}}$}, \scalebox{0.9}{${\tilde{L}}$}, \scalebox{0.9}{${\tilde{Q}}\left( t \right)$} and \scalebox{0.9}{${\tilde{\Phi }}\left( t \right)$} are dual capacitance, dual inductance, dual electric charge, and dual magnetic flux, respectively. The commutation relationship between the dual charge \scalebox{0.9}{${\tilde{Q}}\left(t\right)$} and dual flux \scalebox{0.9}{${\tilde{\Phi }}\left(t\right)$} is:
\begin{align}\label{eq5}
\scalebox{0.85}{$\displaystyle
\left[ \tilde{\Phi }\left( t \right),\tilde{Q}\left(t \right) \right]=i\hbar
$}.
\end{align}
From the dual Hamiltonian of Eq.(\ref{eq4}), the equations of motion are given by:
\begin{align}\label{eq6}
\scalebox{0.85}{$\displaystyle
\tilde{I}\left( t \right)\equiv \frac{d\tilde{Q}\left( t \right)}{dt}=\frac{i}{\hbar }\left[ {{{\tilde{H}}}_{LC}},\tilde{Q}\left( t \right) \right]=-\frac{\tilde{\Phi }\left( t \right)}{{\tilde{L}}}
$},\nonumber \\
\scalebox{0.85}{$\displaystyle
\tilde{V}\left( t \right)\equiv \frac{d\tilde{\Phi }\left( t \right)}{dt}=\frac{i}{\hbar }\left[ {{{\tilde{H}}}_{LC}},\tilde{\Phi }\left( t \right) \right]=\frac{\tilde{Q}\left( t \right)}{{\tilde{C}}}
$},\;\;
\end{align}
where, \scalebox{0.9}{$\tilde{I}\left( t \right)$} and \scalebox{0.9}{$\tilde{V}\left( t \right)$} denoted by the tilde, are the dual current and the dual voltage in the dual quantum $L\!C$ circuit, respectively. As the first step of the dual Hamiltonian method, two dual conditions between equation Eq.(\ref{eq3}) and the dual equations of Eq (\ref{eq6}) are assumed as follows: 
\begin{align}\label{eq7}
\scalebox{0.85}{$\displaystyle
V\left( t \right)\equiv \tilde{I}\left( t \right),\;\;\;                         
I\left( t \right)\equiv \tilde{V}\left( t \right)
$}. 
\end{align} 
The two conditions of Eq .(\ref{eq7})  are called dual conditions. The next step of the dual Hamiltonian method is to derive a relational expression for the canonically conjugate operators that act on each other according to the duality condition (\ref{eq7}).  According to this, the following two relational expressions between charge and flux in dual systems are derived as shown below:  
\begin{align}\label{eq8}
\scalebox{0.85}{$\displaystyle
\Phi \left( t \right)\equiv -\tilde{Q}\left( t \right),\;\;\;         
\tilde{\Phi }\left( t \right)\equiv Q\left( t \right)
$},
\end{align} 
The last step of the dual Hamiltonian method is to derive a relational expression between the constants according to the duality condition of Eq.(\ref{eq7}). According to this, as shown below, two relational expressions between the electrostatic capacitance and the inductance within the dual systems are derived:
\begin{align}\label{eq9}
\scalebox{0.85}{$\displaystyle
           \tilde{C}\equiv L,\quad
           \tilde{L}\equiv -C
$},
\end{align} 
In this section, the conditions under which the dual Hamiltonian of a quantum $L\!C$ circuit, which is a trivial self-dual system, were established. In particular, the duality condition of (\ref{eq7})  is very important, because it becomes an index for defining an exact dual system.
\section{DH method between the JJ and QPSJ in a single junction \label{sec2}}
In this section, according to Hamiltonian of $Q\!P\!S\!J$ introduced by Mooij et al \cite{ref11}-\cite{ref13} which is already known prior research, using the method introduced in the previous section, we investigate the duality between $J\!J$ and $Q\!P\!S\!J$ for case of single junction \cite{ref20}-\cite{ref22}. First, the Hamiltonian \scalebox{0.9}{$H$} of the single $J\!J$ and the Hamiltonian  \scalebox{0.9}{$\tilde{H}$} of the single $Q\!P\!S\!J$ are shown as follows: 
\begin{align}\label{eq10}
\scalebox{0.85}{$\displaystyle
H\Bigl( \theta ,N \Bigr)=E_{c}^{{}}{{N}^{2}}+{{E}_{J}}\Bigl( 1-\cos \theta  \Bigr)
$},\;     
\end{align} 
\begin{align}\label{eq11}
\scalebox{0.85}{$\displaystyle
\tilde{H}\left( \tilde{\theta },\tilde{N} \right)={{E}_{L}}{{\tilde{N}}^{2}}+{{E}_{S}}\left( 1-\cos \tilde{\theta } \right)
$}.
\end{align} 
In Eq.(\ref{eq10}), \scalebox{0.9}{${{E}_{c}}\!\equiv\!{\left(2e\right)\!^2}\!/{2C}$} is charging energy per Cooper pair, therefore, \scalebox{0.9}{${E_J}\!\equiv\!{{{\Phi }_0}{I_c}}/{2\pi }$} is the Josephson energy, \scalebox{0.9}{${I_c}$} and \scalebox{0.9}{${\Phi }_0\!\equiv\!h/2e$} are the critical current and the magnetic flux-quantum, respectively, and \scalebox{0.9}{$N$} and \scalebox{0.9}{$\theta$} are the number of the Cooper pair and the phase of the Cooper pair, respectively. In Eq.(\ref{eq10}),  \scalebox{0.9}{${E_L}\!\equiv\!{{{\Phi }_0}\!^2}\!/2L$} is the inductive energy per magnetic flux quantum, \scalebox{0.9}{${E_S}\!\equiv\!{2e{V_c}}/{2\pi}$} is the $Q\!P\!S$ amplitude, \scalebox{0.9}{${V_c}$} is the critical voltage, \scalebox{0.9}{$\tilde{N}$} and  \scalebox{0.9}{$\tilde{\theta }$} are the number of magnetic flux-quantum and the phase of magnetic flux-quantum in $Q\!P\!S$ junction respectively. The commutation relations by Hamiltonian  \scalebox{0.9}{$H$} and  \scalebox{0.9}{$\tilde{H}$} canonical conjugate variables are described as follows:
\begin{align}\label{eq12}
\scalebox{0.85}{$\displaystyle
\Bigl[ \theta \left( t \right),N\left( {t} \right) \Bigr]=i,\;\;  	
\left[ \tilde{\theta }\left( t \right),\tilde{N}\left( {t} \right) \right]=i
$},
\end{align} 
From the equation of motion for each Hamiltonian, we derived the Josephson's equation for two sets is derived. One set are the usual Josephson's equations as follows:  
\begin{align}\label{eq13}
\scalebox{0.85}{$\displaystyle
V=\frac{\hbar }{2e}\frac{\partial \theta }{\partial t}=\frac{1}{2e}\frac{\partial H}{\partial N}=\frac{2N}{2e}{{E}_{c}}
$}\qquad\qquad\nonumber\\  
\scalebox{0.85}{$\displaystyle
I=2e\frac{\partial N}{\partial t}=\frac{-2e}{\hbar }\frac{\partial H}{\partial \theta }=-\frac{2\pi }{{{\Phi }_{0}}}{{E}_{J}}\sin \theta
$},\;\;
\end{align} 
where \scalebox{0.9}{$I$} and \scalebox{0.9}{$V$} are the current and voltage in the $J\!J$, respectively. The other set are the dual Josephson's equations in the $Q\!P\!S\!J$ as follows \cite{ref20}-\cite{ref22}: 
\begin{align}\label{eq14}
\scalebox{0.85}{$\displaystyle
\tilde{V}=\frac{\hbar }{{{\Phi }_{0}}}\frac{\partial \tilde{\theta }}{\partial t}=\frac{1}{{{\Phi }_{0}}}\frac{\partial \tilde{H}}{\partial \tilde{N}}=\frac{2\tilde{N}}{{{\Phi }_{0}}}{{E}_{L}},\qquad\;\;\nonumber 
$}\\
\scalebox{0.85}{$\displaystyle
\tilde{I}=-{{\Phi }_{0}}\frac{\partial \tilde{N}}{\partial t}=\frac{{{\Phi }_{0}}}{\hbar }\frac{\partial \tilde{H}}{\partial \tilde{\theta }}=\frac{2\pi }{2e}{{E}_{S}}\sin \tilde{\theta }
$}.                    
\end{align}
When the condition of Eq.(\ref{eq7}) is imposed between Eq.(\ref{eq13}) and (\ref{eq14}),  the following two relational expressions between phase and number of particles between dual systems are derived as shown below. One of them is the relationship between the phase  \scalebox{0.95}{$\theta \left( t \right)$} of the Cooper pair and the number \scalebox{0.95}{$\tilde{N}\left( t \right)$} of the magnetic flux-quantum, and the other is the relationship between the phase $\tilde{\theta }\left( t \right)$ of the magnetic flux-quantum and the number \scalebox{0.9}{$N\left( t \right)$} of the Cooper pair \cite{ref20}-\cite{ref22}, as follows: 
\begin{align}\label{eq15}
\scalebox{0.85}{$\displaystyle
\theta \left( t \right)={{\sin }^{-1}}\left[-2\pi \tilde{N}\left( t \right) \right],\quad\;
\tilde{\theta }\left( t \right)={{\sin }^{-1}}\Bigl[2\pi N\left( t \right) \Bigr]
$}.                      
\end{align}
If it is recognized that the relationships described in Eq.(\ref{eq15}) are satisfied, the relationship between the QPS amplitude and charging energy per single-charge, and the relationship between Josephson energy and inductive energy per magnetic flux-quantum, are as follows:
\begin{align}\label{eq16}
\scalebox{0.85}{$\displaystyle
{{E}_{S}}=\frac{1}{2{{\pi}^2}}{E_c},\quad\;
{{E}_{J}}=\frac{1}{2{{\pi}^2}}{E_L}  
$}.      
\end{align}
Furthermore, inductance and capacitance are related to the critical current and the critical voltage, respectively, as follows:
\begin{align}\label{eq17}
\scalebox{0.85}{$\displaystyle
L=\frac{{{\Phi }_0}}{2\pi {I_c}},\quad\;
C=\frac{2e}{2\pi {V_c}}  
$}.             
\end{align}
The linear approximation of Eq.(\ref{eq15}) is a well-known relationship between the phase and the number of particles, as shown in the following equations:
\begin{align}\label{eq18}
\scalebox{0.85}{$\displaystyle
\theta\left( t \right)=-2\pi \tilde{N}\left( t \right)=-2\pi \frac{\Phi }{{{\Phi }_{0}}},\nonumber
$} \\
\scalebox{0.85}{$\displaystyle              
\tilde{\theta }\left( t \right)=2\pi N\left( t \right)=2\pi \frac{Q}{2e}  
$}.\quad\;  
\end{align}
To compare with the existing theoretical formula, calculating the kinetic inductance \scalebox{0.9}{${L_{kin}}$} and the kinematic capacitance \scalebox{0.9}{${C_{kin}}$} defined by \scalebox{0.9}{${L_{kin}}\!^{-1}\!\equiv\!{{{\partial }^2}H}/{\partial {{\Phi }^2}}$} and \scalebox{0.9}{${C_{kin}}\!^{-1}\!\equiv\!{{{\partial }^2}H}/{\partial {Q^2}}$}, respectively, according to Eq.(\ref{eq15}) yields the following equations:
\begin{align}\label{eq19}
\scalebox{0.85}{$\displaystyle
{{L}_{kin}}=\left[ 1-{{\left( 2\pi {\Phi }/{{{\Phi }_{0}}}\; \right)}^{2}} \right]\frac{{{\Phi }_{0}}}{2\pi {{I}_{c}}\cos \theta },\nonumber
$} \\
\scalebox{0.85}{$\displaystyle
{{C}_{kin}}=\left[ 1-{{\left( 2\pi {q}/{2e}\; \right)}^{2}} \right]\frac{2e}{2\pi {{V}_{c}}\cos \tilde{\theta }}\; 
$}.               
\end{align}
To summarize the results of this section, by accepting the results of Eq.(\ref{eq15}) to (\ref{eq17}) obtained under the double condition of Eq.(\ref{eq7}), the Hamiltonian in Eq.(\ref{eq10}) and (\ref{eq11}), it was found that its duality was completely guaranteed.Among the results, the Eq.(\ref{eq15}) is particularly important as it becomes the starting point as a relational expression for creating a self dual system in the next section. The phase and number of particles between the dual systems of Eq.(\ref{eq15}) are nonlinear, and its linear approximation Eq.(\ref{eq18}) is consistent with the generally well known relationship of Mooij et al \cite{ref11}-\cite{ref13}. Eq.(\ref{eq16}),  we note that between the $Q\!P\!S$ amplitude in $Q\!P\!S$ and the charging energy in $J\!J$, between the Josephson energy in  $J\!J$ and the induced energy in $Q\!P\!S$ are connected by \scalebox{0.9}{${1/2{\pi}^2}$} times relation. This relation is important, but it is not mentioned in the paper by Mooij et al \cite{ref11}-\cite{ref13}. The kinetic inductance and kinematic capacitance of Eq.(\ref{eq19}) have the differece with  \scalebox{0.9}{${-2\pi {{\Phi}^2}}\!/\!{{{\Phi }_0}{I_c}\cos \theta }$} and \scalebox{0.9}{${-2\pi {q^2}}\!/\!{2e{V_c}\cos\tilde{\theta}}$} respectively, from the calculation result of Mooij et al \cite{ref11}-\cite{ref13}. 
\section{Simple proof of self-duality in various quantum junction circuits \label{sec3}}
In this section, a simple proof of duality in the single $J\!J$ system is presented. First, the half angle version of Eq.(\ref{eq15}) is introduced as follows: 
\begin{align}\label{eq20}
\scalebox{0.85}{$\displaystyle
\quad\theta \left( t \right)\!=\!2{{\sin }^{-1}}\!\left[\!-\pi \tilde{N}\left( t \right) \right],\quad
\tilde{\theta }\left( t \right)\!=\!2{{\sin }^{-1}}\!\Bigl[\pi N\left( t \right) \Bigr]
$},
\end{align}                  　
Eq.(\ref{eq20}) is in agreement with Eq.(\ref{eq15}) and (\ref{eq18}) within the range of the linear approximation. Substituting Eq.(\ref{eq15}) into the second term of Eq.(\ref{eq10}) and (\ref{eq11}), their Hamiltonians are described as follows:
\begin{align}\label{eq21}
\scalebox{0.85}{$\displaystyle
H\left( \tilde{N},N \right)={{E}_{c}}{{N}^{2}}+2{{\pi }^{2}}{{E}_{J}}{{\tilde{N}}^{2}}
$},
\end{align} 
\begin{align}\label{eq22}
\scalebox{0.85}{$\displaystyle
\tilde{H}\left( N,\tilde{N} \right)={{E}_{L}}{{\tilde{N}}^{2}}+2{{\pi }^{2}}{{E}_{S}}{{N}^{2}}
$}.
\end{align} 
It is trivial that these two Hamiltonians are equal by applying the relation of Eq.(\ref{eq16}) to (\ref{eq21}) and (\ref{eq22}). It is also trivial that Eq.(\ref{eq21}) is quite equivalent to the quantum $L\!C$ circuit discussed in Section 1. Further, when the linear relational expression of Eq.(\ref{eq18}) is used for the second terms of Eq.(\ref{eq21}) and (\ref{eq22}), respectively, the equations are expressed as follows:
\begin{align}\label{eq23}
\scalebox{0.85}{$\displaystyle
H\left( \tilde{N},\theta  \right)={{E}_{c}}{{N}^{2}}+\frac{1}{2}{{E}_{J}}{{\theta }^{2}}
$},
\end{align} 
\begin{align}\label{eq24}
\scalebox{0.85}{$\displaystyle
\tilde{H}\left( \tilde{N},\tilde{\theta } \right)={{E}_{L}}{{\tilde{N}}^{2}}+\frac{1}{2}{{E}_{S}}{{\tilde{\theta }}^{2}}
$}.
\end{align} 
Regarding the second term of Eq.(\ref{eq23}) and (\ref{eq24}), it is obvious that this is a Gaussian approximation of the cosine term of the second term of Eq.(\ref{eq10}) and (\ref{eq11}). Next, substituting Eq.(\ref{eq15}) into the first terms of Eq.(\ref{eq10}) and (\ref{eq11}), their Hamiltonians can be expressed as follows:
\begin{align}\label{eq25}
\scalebox{0.85}{$\displaystyle
H\left( \theta ,\tilde{\theta } \right)=\frac{{{E}_{c}}}{2{{\pi }^{2}}}\left( 1-\cos \tilde{\theta } \right)+{{E}_{J}}\Bigl( 1-\cos \theta  \Bigr)
$},
\end{align} 
\begin{align}\label{eq26}
\scalebox{0.85}{$\displaystyle
\tilde{H}\left( \tilde{\theta },\theta \right)=\frac{{{E}_{L}}}{2{{\pi }^{2}}}\Bigl( 1-\cos \theta  \Bigr)+{{E}_{S}}\left( 1-\cos \tilde{\theta } \right)
$},
\end{align} 
As with the relation of Eq.(\ref{eq21}) and (\ref{eq22}),it is trivial that these two Hamiltonians are equal by substituting Eq.(\ref{eq16}) into (\ref{eq25}) and (\ref{eq26}). Further, when the linear relational expression of Eq.(\ref{eq18}) is substituted in the first terms of Eq.(\ref{eq25}) and (\ref{eq26}), the equations can be rewritten as follows:
\begin{align}\label{eq27}
\scalebox{0.85}{$\displaystyle
  H\Bigl( \theta ,N \Bigr)=\frac{{{E}_{c}}}{2{{\pi }^{2}}}\Bigl[ 1-\cos \left( 2\pi N \right) \Bigr]+{{E}_{J}}\Bigl( 1-\cos \theta  \Bigr)
  $},
\end{align} 
\begin{align}\label{eq28}
\scalebox{0.85}{$\displaystyle
\tilde{H}\left( \tilde{\theta },\tilde{N} \right)=\frac{{{E}_{L}}}{2{{\pi }^{2}}}\left[ 1-\cos ( 2\pi \tilde{N} ) \right]+{{E}_{S}}\left( 1-\cos \tilde{\theta } \right)
$}. 
\end{align} 
By imposing the conditions of Eq.(\ref{eq17}) and (\ref{eq18}) on Eq.(\ref{eq27}) and (\ref{eq28}), these two Hamiltonians are equal, i.e. self-dual. A Hamiltonian with two cosine terms, competing with each other, similar to Eq.(\ref{eq25}) to (\ref{eq28}) is a new form which has not been known until now. Such a system is a system in which both $J\!J$ and  $Q\!P\!S\!J$ which are in a coherent state compete with each other, and the circuit in which $J\!J$ and $Q\!P\!S\!J$ are connected in series is called a $J\!J$- $Q\!P\!S\!J$ competitive circuit. FIG.1 shows the equivalent circuits for various self-dual systems. FIG.1 (a) shows the quantum $L\!C$ circuit represented by the Hamiltonian of Eq.(\ref{eq21}) and (\ref{eq22}) or Eq.(\ref{eq23}) and (\ref{eq24}). FIG.1 (b) and (c) show a single $J\!J$ represented by the Hamiltonian of Eq.(\ref{eq10}) and a single $Q\!P\!S\!J$ represented by the Hamiltonian of the Eq.(\ref{eq11}), respectively. FIG.1 (d) shows the $J\!J$- $Q\!P\!S\!J$ competitive circuit represented by the Hamiltonian of Eq.(\ref{eq27}) and (\ref{eq28}).

\begin{figure}[hbt]
\includegraphics[width=0.95\columnwidth]{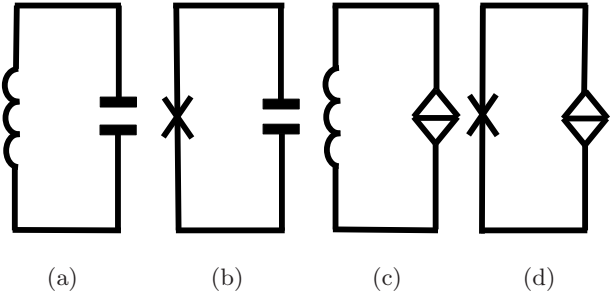}\\
\begin{tabular}{cccc}
\;(a) &\qquad\qquad\quad(b) &\qquad\quad\quad\;\;\;(c) & \qquad\quad\quad\;\:(d)\;\;\\
\end{tabular}
\caption{Equivalent circuits for various self-dual systems. (a) quantum LC circuit. (b) JJ. (c) QPS junction. (d) JJ - QPSJ competitive circuit..}
\end{figure}
\section{Superconductor- insulator transitions \label{sec4}}
In this section, superconductor–insulator transition \cite{ref23}-\cite{ref25} are discussed from the viewpoint of quantum phase transition. Quantum resistance was derived using the following two methods with the Josephson's equations of Eq.(\ref{eq13}) and the dual Josephson's equations of Eq.(\ref{eq14}) and (\ref{eq15}). One of the methods uses the ratio between the fluctuation of the number of Cooper pairs and the fluctuation of the number of magnetic flux-quantums. The resistance can thus be derived as follows: 
\begin{align}\label{eq29} 
\scalebox{0.85}{$\displaystyle
R=\frac{V}{I}=\frac{R_Q}{2\pi }\frac{d\theta }{dN}=\frac{R_Q}{2{{\pi }^2}}\frac{E_c}{E_J}\frac{N}{{\tilde{N}}}
$},
\end{align} 
where \scalebox{0.9}{${R_Q}\!\equiv\!{h}/{{{\left( 2e \right)}^2}}\!\approx\! 6.4\left[ k\Omega  \right]$} is the universal critical sheet resistance. The other method uses the ratio between the fluctuation of the phase of the Cooper pair and the fluctuation of the phase of the magnetic flux-quantum. The quantum conductance can thus be derived as follows: 
\begin{align}\label{eq30} 
\scalebox{0.85}{$\displaystyle
G=\frac{\tilde{V}}{\tilde{I}}=-\frac{1}{2\pi }{G_Q}\frac{d\tilde{\theta }}
{d\tilde{N}}={G_Q}\frac{1}{2{{\pi }^2}}\frac{E_L}{E_S}\frac{\tilde{N}}{N}
$},
\end{align} 
where \scalebox{0.9}{${G_Q}\equiv {R_Q}^{-1}$} is the universal critical sheet conductance. In Eq.(\ref{eq29}) when the conditions \scalebox{0.9}{$\Delta \theta\!>>\!\Delta N$}, or \scalebox{0.9}{${E_c}\!>>\!{E_J}$} and \scalebox{0.9}{$N\!>>\!\tilde{N}$}, are met, the equation represents an insulator state. In particular, when R→∞, it represents a superinsulator state \cite{ref20}, \cite{ref26}-\cite{ref28}. The reverse case occurs when the conditions of \scalebox{0.9}{$\Delta \theta\!<<\!\Delta N$} or, \scalebox{0.9}{${E_c}\!<<\!{E_J}$} and \scalebox{0.9}{$N\!<<\!\tilde{N}$}, are met, the equation represents a conductor state. In particular, when $R$→0, it represents a superconductor state. In the special case of  \scalebox{0.9}{$\Delta \theta \approx 2\pi \Delta N$}, or \scalebox{0.9}{${E_c}\approx 2{{\pi }^{2}}{E_J}$} and \scalebox{0.9}{$N\simeq \tilde{N}$}, the equation represents a critical state. In Eq.(\ref{eq30}), when the conditions: \scalebox{0.9}{$\Delta \tilde{\theta }\!>>\!\Delta \tilde{N}$}, or \scalebox{0.9}{${E_L}\!>>\!{E_S}$} and \scalebox{0.9}{$\tilde{N}\!>>\!N$}, are met, the equation represents a conductor state. In particular, when $G$→∞, it represents a superconductor state. The reverse case occurs when the conditions of \scalebox{0.9}{$\Delta \tilde{\theta }\!<<\!\Delta \tilde{N}$}, or \scalebox{0.9}{${E_L}\!<<\!{{E}_{S}}$} and \scalebox{0.9}{$\tilde{N}\!<<\!N$}, are met, in which case the equation represents an insulator state. In particular, when $G$→0, it represents a superinsulator state.

\section{Partition function of the $J\!J$ and $Q\!P\!S\!J$ in single junction \label{sec5}}
Up to the preceding section, the $J\!J$ and $Q\!P\!S\!J$ have been dealt with in a single junction for Hamiltonian form at the level of classical mechanics. In this section, the partition function of a single junction is investigated, incorporating the quantum effect by path integral and its duality is considered \cite{ref20}. The partition function of a single $J\!J$ system is expressed using Eq.(\ref{eq10}) as follows:  
\begin{align}\label{eq31}
\scalebox{0.85}{$\displaystyle
Z\!\!=\!\!\int\!\!{DN}D\theta \exp \frac{1}{\hbar }\!\!\int\limits_{0}^{\beta }\!\!{d\tau }\!\left[ i\hbar \frac{\partial \theta }{\partial \tau }N\!-\!{{E}_{c}}{{N}^{2}}\!-\!{{E}_{J}}\Bigl(1\!-\!\cos \theta  \Bigr) \right]$},
\end{align}
where, \scalebox{0.82}{$\int\!{DN}\!\equiv\!\prod\limits_{\tau }{\int_{-\infty}^{\infty}\!dN(\tau)}$}, \scalebox{0.82}{$\int\!{D\theta }\!\equiv\!\prod\limits_{\tau }{\int_{-\pi }^{\pi }\!{d\theta(\tau)}/{2\pi}}$}, \scalebox{0.82}{$\beta\!\equiv\!{({k_B}T)^{-1}}$}, and \scalebox{0.98}{$\tau\!\equiv\!\beta \hbar$} is the imaginary time. In the same manner, the partition function of a single $Q\!P\!S\!J$ system is expressed using Eq.(\ref{eq11}) as follows: 
\begin{align}\label{eq32}
\scalebox{0.85}{$\displaystyle
\tilde{Z}\!\!=\!\!\int\!\!\!{D\tilde{N}}D\tilde{\theta }\exp\frac{1}{\hbar}\!\!\int\limits_{0}^{\beta }\!\!{d\tau }\!\!\left[i\hbar\frac{\partial \tilde{\theta }}{\partial \tau }\tilde{N}\!-\!{{E}_{L}}{{\tilde{N}}^{2}}\!-\!{{E}_{S}}\left( 1\!-\!\cos\tilde{\theta}\right) \right]$}.
\end{align} 
First, to make computation using path integration of Eq.(\ref{eq31}) and (\ref{eq32}) convenient, imaginary time $\tau$ is changed from the continuous value to the discrete value, the differential operator is changed to the difference operator, and the integral $\int_{0}^{\beta }{d\tau }$ is changed to the sum $\sum\nolimits_{\tau =1}^{M_{\tau }}$. These partition functions can then be expressed as follows:
\begin{align}\label{eq33}
\scalebox{0.85}{$\displaystyle 
Z\!\!=\!\!\int{\!\!DN}\!\!D\theta\exp\sum\limits_{\tau =1}^{{M_{\tau }}}{\left[ iN{{\nabla }_{\tau }}\theta\!-\!{E'_c}{N^2}\!-\!{E'_J}\Bigl(1\!-\!\cos \theta  \Bigr) \right]} $},
\end{align}
\begin{align}\label{eq34}
\scalebox{0.85}{$\displaystyle 
\tilde{Z}\!\!=\!\!\int{\!\!D\tilde{N}}\!\!D\tilde{\theta }\exp\sum\limits_{\tau =1}^{{M_{\tau }}}{\left[ i\tilde{N}{{\nabla }_{\tau }}\tilde{\theta }\!-\!{E'_L}{{\tilde{N}}^2}\!-\!{E'_S}\left( 1\!-\!\cos \tilde{\theta } \right) \right]} $},
\end{align}
here \scalebox{0.9}{$E'\!_c$}, \scalebox{0.9}{$E'\!_J$}, \scalebox{0.9}{$E'\!_L$} and \scalebox{0.9}{$E'\!_S$} are the dimensionless energy defined by \scalebox{0.9}{${\Delta\tau{E\!_c}}/{\hbar }$}, \scalebox{0.9}{${\Delta\tau{E\!_J}}/{\hbar}$}, \scalebox{0.9}{${\Delta \tau{E\!_L}}/{\hbar}$} and \scalebox{0.9}{${\Delta\tau {E\!_S}}/{\hbar}$}, respectively. In addition, \scalebox{0.9}{$\Delta\tau\!\!\equiv\!\!{{\tau}_{\max }}/{M_{\tau}}$}, \scalebox{0.9}{${{\tau}_{\max}}$, ${M_{\tau}}$} and \scalebox{0.9}{${{\nabla\!}_{\tau }}\!\theta(\tau)\!\!\equiv\!\!\theta(\tau)\!\!-\!\!\theta(\!\tau\!\!-\!\!\Delta\tau\!)$} are the minimum imaginary time interval, the maximum imaginary time, the division number and the difference operator in imaginary time, respectively. When Eq.(\ref{eq33}) and (\ref{eq34}) are integrated with respect to \scalebox{0.9}{$N\!\left( \tau  \right)$} and \scalebox{0.9}{$\tilde{N}\!\left( \tau  \right)$}, respectively, the following equations are obtained: 
\begin{align}\label{eq35}
\scalebox{0.85}{$\displaystyle 
Z\!\!=\!\!\int{\!\!D\theta }\exp\!\sum\limits_{\tau =1}^{{M_{\tau }}}{\left[ -\frac{1}{2}E{{_J^0}^{\prime }}{{\Bigl( {{\nabla }_{\tau }}\theta \Bigr)}^2}\!-\!{E'_J}\Bigl( 1-\cos \theta  \Bigr) \right]}
$},          
\end{align}
\begin{align}\label{eq36}
\scalebox{0.85}{$\displaystyle 
\tilde{Z}\!\!=\!\!\int{\!\!D\tilde{\theta }}\exp\!\!\sum\limits_{\tau =1}^{{M_{\tau }}}{\left[ -\frac{1}{2}E{{_S^0}^{\prime }}{{\left( {{\nabla }_{\tau }}\tilde{\theta } \right)}^2}\!-\!{E'_S}\left( 1-\cos \tilde{\theta } \right) \right]}
$},
\end{align}
where \scalebox{0.9}{$E{_J^0}^{\prime }$} and \scalebox{0.9}{$E{_S^0}^{\prime }$} represent the dimensionless energy of the imaginary time component in the $J\!J$ and $Q\!P\!S\!J$, respectively, and are defined by the following equations: 
\begin{align}\label{eq37}
\scalebox{0.95}{$\displaystyle
{E{_J^0}^{\prime }}\equiv\frac{1}{2{E'_c}}
$},
\end{align}
\begin{align}\label{eq38}
\scalebox{0.95}{$\displaystyle
{E{_S^0}^{\prime }}\equiv\frac{1}{2{E'_L}}
$},
\end{align}
On the contrary, when Eq.(\ref{eq33}) and (\ref{eq34}) are integrated with respect to \scalebox{0.9}{$\theta(\tau)$} and \scalebox{0.9}{$\tilde{\theta}(\tau)$}, respectively, the following equations are obtained \cite{ref20}: 
\begin{align}\label{eq39}
\scalebox{0.98}{$\displaystyle
Z\!\!=\!\!\int{\!\!D{N}}\!\exp\!\sum\limits_{\tau =1}^{{M_{\tau }}}\!{\Bigl[ \!-\!{E}'_J\!-\!{E}'_cN^2
\!+\!\ln{I_{\alpha\left(\tau\right)}}\!{\left({{E'}_{\!\!J}} \right)} 
\Bigr]}
$},
\end{align}
\begin{align}\label{eq40}
\scalebox{0.98}{$\displaystyle
\tilde{Z}\!\!=\!\!\int{\!\!D\tilde{N}}\!\exp\! \sum\limits_{\tau =1}^{{{M}_{\tau }}}\!{\left[\!-\!{E}'_S\!-\!{E}'_L\tilde{N}^2
\!\!+\!\ln {I_{\tilde{\alpha }\left(\tau\right)}}\!{\left({{E}'_S} \right)}
\right]}
$},
\end{align}
where \scalebox{0.9}{${I_{\alpha\left(\tau\right)}}\!\!\left({E'_J}\right)$} and \scalebox{0.9}{${I_{\tilde{\alpha}\left(\tau\right)}}\!\!\left( {E'_S} \right)$} represent modified Bessel functions of order \scalebox{0.9}{$\alpha\!\left(\tau\right)\!\!\equiv\!\!-\!{\nabla\!_\tau}N\!\!\left(\tau\right)$} and order \scalebox{0.9}{$\tilde{\alpha}\!\left(\tau\right)\!\!\equiv\!\!-\!{\nabla\!_\tau}\tilde{N}\!\!\left(\tau\right)$} respectively, When the Villain approximation \cite{ref29}-\cite{ref30} is introduced into the modified Bessel functions of Eq.(\ref{eq39}) and (\ref{eq40}), the following equations are obtained:
\begin{align}\label{eq41}
\scalebox{0.82}{$\displaystyle
Z\!\!=\!\!\int{\!\!D{N}}\!\exp\!\sum\limits_{\tau =1}^{{M_{\tau }}}\!\!{\left[\!-\!{E}'_J\!-\!{E}'_cN^2
\!+\!\ln {I_0}{\left({{E'}_{\!\!J}} \right)}
\!-\!\frac{1}{2{{\left(E'_J\right)}_v}}{{\!\Bigl({\nabla_\tau}N\Bigr)\!}^2}        
\right]}
$},
\end{align}
\begin{align}\label{eq42}
\scalebox{0.82}{$\displaystyle
\tilde{Z}\!\!=\!\!\int{\!\!D\tilde{N}}\!\exp\! \sum\limits_{\tau =1}^{{{M}_{\tau }}}\!\!{\left[\!-\!{E}'_S\!-\!{E}'_L\tilde{N}^2
\!\!+\!\ln {I_{0}}\!{\left({{E}'_S} \right)}\!-\!\frac{1}{2{{\left(E'_S\right)}_v}}{{\left({\nabla_\tau}\tilde{N}\right)\!}^2}
\right]}
$},
\end{align}
where \scalebox{0.9}{${\left(E'_J\right)_v}$} and \scalebox{0.9}{${\left(E'_S\right)_v}$} are Villain's parameters \cite{ref30}-\cite{ref31} and are defined as follows: 
\begin{align}\label{eq43}
\scalebox{0.80}{$\displaystyle
{\left(E'_J\right)_v}\equiv \frac{-1}{2}\frac{1}{\ln \Bigl[{I_1\left({E'_J}\right)}/{{I_0}\left({E'_J} \right)}\; \Bigr]}
$},\quad
\end{align}
\begin{align}\label{eq44}
\scalebox{0.80}{$\displaystyle
{\left(E'_S\right)_v}\equiv \frac{-1}{2}\frac{1}{\ln \Bigl[ {I_1\left({E'_S}\right)}/{{I_0}\left({E'_S} \right)}\; \Bigr]}
$},\quad
\end{align}
When the conversion formula between the number of particles and its dual phase in Eq.(\ref{eq15}) is substituted in Eq.(\ref{eq41}) and (\ref{eq42}), the following equations are obtained \cite{ref20}:
\begin{align}\label{eq45}
\scalebox{0.85}{$\displaystyle
 Z\!\!\approx\!\!\int{\!\!D\tilde{\theta }}\exp \sum\limits_{\tau =1}^{M_{\tau }}{\left[ -{E'_J}+\ln \Bigl( \frac{1}{2\pi }\cos\frac{\tilde{\theta }}{2} \Bigr) +\ln{I_0}\left(E'_J\right)
 \right.}$}\nonumber\\
\scalebox{0.85}{$\displaystyle
\left.
-\frac{{{\cos }^{2}}({{\tilde{\theta }}}/{2})}{8{{\pi }^{2}}{{\left(E'_J\right)}_v}}{{\left( {{\nabla }_{\tau }}\tilde{\theta }\right)}^2}-\frac{{{E'}_c}}{2{{\pi }^{2}}}\left(1-\cos \tilde{\theta }\right) \right]
 $},\quad
\end{align}
\begin{align}\label{eq46}
\scalebox{0.85}{$\displaystyle
\tilde{Z\!\!}\approx\!\!\int{\!\!D\theta}\exp \sum\limits_{\tau =1}^{M_{\tau }}{\left[ -{E'_S}+\ln \Bigl( \frac{-1}{2\pi} \cos \frac{\theta }{2}\Bigr)+\ln {I_0}\left(E'_S\right)
\right.}$}\nonumber\\
\scalebox{0.85}{$\displaystyle
\left.
-\frac{{{\cos }^{2}}\left( {\theta }/{2}\; \right)}{8{{\pi }^{2}}{{\left(E'_S\right)}_v}}{{\Bigl( {{\nabla }_{\tau }}\theta\Bigr)}^2}-\frac{{{E}_{L}}^{\prime }}{2{{\pi }^{2}}}\Bigl(1-\cos \theta\Bigr) \right]
 $}.\quad
\end{align}
By comparing Eq.(\ref{eq45}) with (\ref{eq36}), and by considering that \scalebox{0.9}{$\left(E'_J\right)_v\!\approx\! E'_J$} and \scalebox{0.9}{${\cos }^2(\tilde{\theta }/2)\!\approx\!1$} are established respectively at the limit of large \scalebox{0.9}{${E'_J}$} and the limit of small \scalebox{0.9}{$\tilde{\theta }$}, it can be understood that the relational expressions \scalebox{0.9}{${E_S}\!=\!E_c/{2{\pi }^2}$} and \scalebox{0.9}{${E_L}\!=\!2{{\pi }^2}{E_J}$}, introduced in Eq.(\ref{eq16}) , are established. Similarly, by comparing Eq.(\ref{eq46}) with (\ref{eq35}), and by considering that \scalebox{0.9}{${\left(E'_S\right)_v}\!\approx\!{E'_S}$} and \scalebox{0.9}{${{\cos }^2}\left( {\theta }/2\right)\!\approx\!1$} are established respectively at the limit of large \scalebox{0.9}{${E'_S}$} and the limit of small \scalebox{0.9}{$\theta$}, it can be understood that the relational expressions of \scalebox{0.9}{${E_J}\!=\!{E_L}/{2{\pi }^2}$} and \scalebox{0.9}{${E_c}\!=\!2{{\pi }^2}{E_S}$} introduced in Eq.(\ref{eq16}) are established. From the above results, at least at the level of the Villain approximation, the partition functions of Eq.(\ref{eq31}) and (\ref{eq32}) are proved to be in a dual relationship with each other.
\section{Partition function of the JJ and QPSJ in a one-dimensional nanowire \label{sec6}}
In the previous sections, the duality for the $J\!J$ and  $Q\!P\!S\!J$ was examined in a single junction. In this section, this is extended to consider the dual model for the $J\!J$ and $Q\!P\!S\!J$ in a nanowire, which is a one-dimensional system. The Hamiltonians obtained by extending Eq.(\ref{eq10}) and (\ref{eq11}) into a one-dimensional nanowire are as follows:
\begin{align}\label{eq47}
\scalebox{0.86}{$\displaystyle
H\left( \theta ,N \right)={{E}_{c}}\sum\limits_{x=i}^{{{M}_{x}}}\left\{{N{{\left( x,\tau  \right)}^{2}}}+{{E}_{J}}{\Bigl[1-\cos {{\nabla }_{x}}\theta \left( x,\tau  \right) \Bigr]}\right\}       
 $},
\end{align}
\begin{align}\label{eq48}
\scalebox{0.86}{$\displaystyle
\tilde{H}( \tilde{\theta },\tilde{N})={{E}_{L}}\sum\limits_{x=i}^{{{M}_{x}}}\left\{{\tilde{N}{{\left( x,\tau  \right)}^{2}}}+{{E}_{S}}{\left[ 1-\cos {{\nabla }_{x}}\tilde{\theta }\left( x,\tau  \right) \right]}\right\}                   
 $},
\end{align}
where \scalebox{0.9}{$x$}, \scalebox{0.9}{$a$}, \scalebox{0.9}{$L$}, \scalebox{0.9}{${M_x}\!\equiv\!{L}/{a}$} and \scalebox{0.9}{${{\nabla }_{x}}\theta\!\left(x,\tau \right)\!\equiv\!\theta\!\left(x,\tau\right)\!-\!\theta\!\left(x\!-\!a,\tau\right)$} are the space variable, the lattice spacing, the length of the one-dimensional nanowire, the division number of the space and difference operator of the space, respectively. The partition functions of Eq.(\ref{eq47}) and (\ref{eq48}) can be expressed as follows:
\begin{align}\label{eq49}
\scalebox{0.90}{$\displaystyle
Z\!\!=\!\!\!\int{\!\!\!D\!N}\!\!D\!\theta
\exp\!\!\sum_{\tau =1}^{M_{\tau }}\!{\sum_{x=1}^{M_x}\!\left[i N{\nabla }_{\tau }\theta
\!\!-\!\!{E'_c}N^2\!\!-\!\!{E'_J} \Bigl(1\!-\!\cos\!{{\nabla\!}_x}\theta\Bigr)\!\right]}
$}
\end{align}
\begin{align}\label{eq50}
\scalebox{0.85}{$\displaystyle
\tilde{Z}\!\!=\!\!\!\int{\!\!\!D\!\tilde{N}}\!\!D\!\tilde{\theta }
\exp\!\!\sum_{\tau =1}^{M_{\tau }}\!{\sum_{x=1}^{M_x}\!\left[i \tilde{N}{\nabla }_{\tau }\tilde{\theta }
\!\!-\!\!{E'_L}\tilde{N}^2\!\!-\!\!{E'_S} \Bigl(1\!-\!\cos\! {{\nabla\! }_x}\tilde{\theta }\Bigr)\!\right]}
$}
\end{align}
where, \scalebox{0.9}{$Z$} and \scalebox{0.9}{$\tilde{Z}$} represent the partition function of the $J\!J$ and the $Q\!P\!S\!J$, respectively, in the one-dimensional nanowire. When Eq.(\ref{eq49}) and (\ref{eq50}) are integrated with respect to \scalebox{0.9}{$N\left( x,\tau  \right)$} and \scalebox{0.9}{$\tilde{N}\left( x,\tau  \right)$}, respectively, the following equations are obtained:
\begin{align}\label{eq51}
\scalebox{0.85}{$\displaystyle
Z\!\!=\!\!\!\int\!\!\!D\theta\!\exp\!\sum_{x,\tau}\left[\!-\frac{1}{2}{{E'_J\!}^0}
\Bigl( {{\nabla }_{\tau }}\theta \Bigr)^2\!\!\!-\!{E'_J}\Bigl(1\!\!-\!\!\cos{{\nabla\!}_x}\theta \Bigr)
\right] 
$},
\end{align}
\begin{align}\label{eq52}
\scalebox{0.85}{$\displaystyle
\tilde{Z}\!\!=\!\!\!\int\!\!\!D\tilde{\theta}\!\exp\!\sum_{x,\tau}\left[\!-\frac{1}{2}{{E'_S\!}^0}
\Bigl( {{\nabla }_{\tau }}\tilde{\theta } \Bigr)^2\!\!\!-\!{E'_S}\Bigl(1\!\!-\!\!\cos{{\nabla\!}_x}\tilde{\theta } \Bigr)
\right]
$},
\end{align}
where, \scalebox{0.9}{$\sum_{x,\tau}\!\equiv\!\sum\nolimits_{\tau\! =\!1}^{M_{\tau }}\sum\nolimits_{x\! =\!1}^{M_x}$}, the first terms of Eq.(\ref{eq51}) and (\ref{eq52})  are expressed in a quadratic form for the imaginary time difference of each phase, but the second terms are expressed in a cosine form for the spatial difference of each phase, but these second terms are expressed in a cosine form for the spatial difference of each phaseHere, in consideration of the periodicity of the lattice space, the cosine form is also introduced for the first terms of the Eq.(\ref{eq51}) and (\ref{eq52}), as follows:
\begin{align}\label{eq53}
\scalebox{0.85}{$\displaystyle
{Z_{\!A\!X\!Y}}\!\!=\!\!\!\!\int\!\!\!D{\theta}\!\exp\!\sum_{x,\tau}\!\left[\!-\!{{E'_J\!}^0}\Bigl(\!1\!-\!\cos{{\nabla\!}_{\tau }}\theta\Bigr)\!\!-\!{E'_J}\Bigl(\!1\!-\!\cos{{\nabla\!}_x}\theta\Bigr)
\right]
$},
\end{align}
\begin{align}\label{eq54}
\scalebox{0.85}{$\displaystyle
{Z_{\!D\!A\!X\!Y}}\!\!=\!\!\!\!\int\!\!\!D\tilde{\theta}\!\exp\!\sum_{x,\tau}\!\left[\!-\!{{E'_S\!}^0}\Bigl(\!1\!-\!\cos{{\nabla\!}_{\tau }}\tilde{\theta } \Bigr)\!\!-\!{E'_S}\Bigl(\!1\!-\!\cos{{\nabla\!}_x}\tilde{\theta}\Bigr)
\right]
$},
\end{align}
where, \scalebox{0.9}{${{Z}_{\!A\!X\!Y}}$} and \scalebox{0.9}{${{Z}_{\!D\!A\!X\!Y}}$} represent the partition function of the anisotropic \scalebox{0.9}{$XY(A\!X\!Y)$} model and the dual anisotropic \scalebox{0.9}{$XY(D\!A\!X\!Y)$} model, respectively, in $1\!+\!1$ dimensions. That is, the \scalebox{0.9}{$A\!X\!Y$} model in the $1\!+\!1$ dimension of Eq.(\ref{eq53})  is equivalent to the $J\!J$ model in the one-dimensional nanowire of Eq.(\ref{eq49}), and the \scalebox{0.9}{$D\!A\!X\!Y$} model in the $1\!+\!1$ dimension of Eq.(\ref{eq54}) is equivalent to the $Q\!P\!S\!J$ model in the one-dimensional nanowire of Eq.(\ref{eq50}). These relationships, which are known as one-dimensional quantum models, are equivalent to $1\!+\!1$ dimensional classical \scalebox{0.9}{$XY$} models \cite{ref32}-\cite{ref34}. In Eq.(\ref{eq53}) and (\ref{eq54}), to make handling convenient, the partition functions \scalebox{0.9}{${Z'_{\!A\!X\!Y}}$} and \scalebox{0.9}{${Z'_{\!D\!A\!X\!Y}}$}, are defined, the constant term is removed and a pure cosine exponent remains as follows:
\begin{align}\label{eq55}
\scalebox{0.85}{$\displaystyle
{Z_{\!A\!X\!Y}}\!\equiv\!\exp\!\!\left[\!-\!\left( {E'}_{J}^0+{E'\!\!}_J \right){{M}_{\tau }}{{M}_{x}} \right]{Z'_{AXY}},
$}\nonumber\\
\scalebox{0.85}{$\displaystyle
{Z'_{\!A\!X\!Y}}\!\!\equiv\!\!\int\!\!\!D{\theta}\exp\!\sum_{x,\tau}\!\!\left(\!
{E'}_{\!\!J}^{0}\cos\! {{\nabla }_{\tau }}\theta\!+\!{E'}_{\!\!J}\cos\! {{\nabla }_{x}}\theta
\right)
$},
\end{align}
\begin{align}\label{eq56}
\scalebox{0.85}{$\displaystyle
{Z_{\!D\!A\!X\!Y}}\!\equiv\!\exp\!\!\left[\!-\!\!\left( {E'}_{\!\!S}^{0}+{E'}_{\!\!S} \right)\!\!{{M}_{\tau }}{{M}_{x}} \right]{Z'_{DAXY}},
$}\nonumber\\
\scalebox{0.85}{$\displaystyle
{Z'_{\!D\!A\!X\!Y}}\!\!\equiv\!\!\int\!\!\!D\tilde{\theta}\exp\!\sum_{x,\tau}\!\!\left(\!
{{E'}_{\!\!S}^{0}}\cos\!{{\nabla\!}_{\!\tau }}\tilde{\theta }\!+\!{E'}_{\!\!S}\cos\!{{\nabla\!}_{\!x}}\tilde{\theta }
\right)
$},
\end{align}
${Z'_{\!A\!X\!Y}}$ and ${Z'_{\!D\!A\!X\!Y}}$ are the starting points for discussing dual transformation by the Villain approximation in the next section. 
\section{Duality between the AXY model and DAXY model by Villain approximation \label{sec7}}
The Villain approximation \cite{ref29}-\cite{ref31} is first applied to ${{{Z}'}_{\!A\!X\!Y}}$ and ${{{Z'}}_{\!D\!A\!X\!Y}}$, introduced in the previous section, as follows:
\begin{align}\label{eq57}
\scalebox{0.67}{$\displaystyle
Z\!_{QV}\!\!\equiv\!\!R_{QV}\!\!\!\int\!\!\!D{\theta}\sum\limits_{\left\{ {n} \right\}}\exp\!\!\sum_{x,\tau}\!\left[\frac{\!\!-\!{\left(\! {E'}_{J}^0 \!\right)\!_v}}{2}{{\Bigl(\!{\nabla }_{\tau }\theta\!-\!2\pi{n_0}\!\Bigr)}\!^2}\!\!
+\!\!\frac{-{{\Bigl(\! {E'_J} \!\Bigr)}\!_v}}{2}{{\Bigl(\! {{\nabla }_{x}}\theta\!-2\pi{{n}_{x}}\!\Bigr)}\!^2} 
\!\right] $},
\end{align}
\begin{align}\label{eq58}
\scalebox{0.67}{$\displaystyle
Z\!_{\!Q\!D\!V}\!\!\equiv\!\!R_{\!Q\!D\!V}\!\!\!\int\!\!\!D\tilde{\theta }\!\sum\limits_{\left\{ {\tilde{n}} \right\}}\!\exp\!\!\sum_{x,\tau}\!\!\left[\frac{\!-\!{\left(\!{E'}_{S}^0\!\right)\!_v}}{2}{\Bigl(\!{\nabla }_{\tau }\tilde{\theta }\!-\!2\pi {\tilde{n}_0}\!\Bigr)}\!^2\!\!
+\!\!\frac{-{{\Bigl(\!{E'_S} \!\Bigr)}\!_v}}{2}{\Bigl(\! {{\nabla }_{x}}\tilde{\theta }\!-2\!\pi{{\tilde{n}}_x}\!\Bigr)}\!^2
\right] $},
\end{align}
where \scalebox{0.9}{$\displaystyle Z_{\!Q\!V}$} and \scalebox{0.9}{$\displaystyle Z_{\!Q\!D\!V}$} are Villain approximations of the partition functions \scalebox{0.8}{$\displaystyle{Z'}_{\!A\!X\!Y}$} and \scalebox{0.8}{$\displaystyle {Z'}_{\!D\!A\!X\!Y}$} respectively, \scalebox{0.8}{$\displaystyle R_{\!Q\!V}\!\!\equiv\!\!\left[R_v\!\left({E'_J}\right)\!R_v\!({E'_J\!}^0)\right]^{{M_x}{M_{\tau }}}\!\!\! $} and \scalebox{0.8}{$\displaystyle R_{\!Q\!D\!V}\!\!\equiv\!\!\left[R_v\!\left({E'_S}\right)\!R_v\!({E'_S\!}^0)\right]^{{M_x}{M_{\tau }}}\!\!\! $} are Villain’s normalization parameters, \scalebox{0.8}{$\displaystyle {R_v\!}\left(E\right)$} is defined as \scalebox{0.8}{${\!R_v\!}\left(E\right)\!\!\equiv\!\!\sqrt{2\pi {{\left(E\right)}_{v}}}{I_0}\left(E\right) $}, The summation symbols \scalebox{0.75}{$\sum\limits_{\left\{{n}\right\}}{\!\equiv\!\!\!\!\sum\limits_{{n_0}\left(\!x,\tau \!\right)\!=\!-\infty}^{\infty}}$}\,\scalebox{0.75}{${\sum\limits_{{n_x}\left(\!x,\tau \!\right)\!=\!-\infty}^{\infty}}$} and \scalebox{0.75}{$\sum\limits_{\left\{{\tilde{n}} \right\}}{\!\equiv\!\!\!\!\sum\limits_{{\tilde{n}_0}\left(\!x,\tau\!\right)\!=\!-\infty}^{\infty}}$}\,\scalebox{0.75}{${\sum\limits_{{{\tilde{n}}_x}\left(\!x,\tau\!\right)\!=\!-\infty}^{\infty}}$} are used for the integer fields \scalebox{0.8}{${n_0}\!\left(\!x,\tau\!\right)$}, \scalebox{0.8}{${n_x}\!\left(\!x,\tau\!\right)$, ${{\tilde{n}}_0}\!\left(\!x,\tau\!\right)$} and \scalebox{0.8}{${{\tilde{n}}_x}\!\left(\!x,\tau\!\right)$} respectively. For Eq.(\ref{eq57}) and (\ref{eq58}) the following identities associated with the Jacobi theta function are used:
\begin{align}\label{eq59}
\scalebox{0.75}{$\displaystyle
\sum_{n=-\infty}^{\infty}\exp\left[-\frac{E}{2}\left(\theta-2\pi n\right)^2\right]=\sum_{b=-\infty}^{\infty}\frac{1}{\sqrt{2\pi E}}\exp\left(-\frac{b^2}{2E}+i b\theta\right)
$},
\end{align}
As a result, Eq.(\ref{eq57}) and (\ref{eq58}) can be rewritten as follows:
\begin{align}\label{eq60}
\scalebox{0.85}{$\displaystyle
Z_{\!Q\!V}\!\!=\!\!{C_{\!Q\!V}}\sum_{\left\{{b}\right\}}{{\delta}_{{\nabla}_j{b_j},0}}\exp\sum_{x,\tau}\!
\left[
\frac{-b_{0}^{2}(x,\tau)}{2{{\left({E'}_{J}^{0}\right)}_v}}+\frac{-b_{x}^{2}(x,\tau)}{2{{\left({{E'}_J} \right)}_v}}
\right]
$},
\end{align}
\begin{align}\label{eq61}
\scalebox{0.85}{$\displaystyle
Z_{\!Q\!D\!V}\!\!=\!\!{C_{\!Q\!D\!V}}\!\!\sum_{\left\{\tilde{b}\right\}}{{\delta}_{{\nabla}_j{\tilde{b}_j},0}}\exp\sum_{x,\tau}\!
\left[
\frac{-\tilde{b}_{0}^{2}(x,\tau)}{2{{\left({E'}_{S}^{0}\right)}_v}}+\frac{-\tilde{b}_{x}^{2}(x,\tau)}{2{{\left({{E'}_S} \right)}_v}}
\right]
$},
\end{align}
where \scalebox{0.9}{$C_{\mathrm{\!Q\!V}}$} and \scalebox{0.9}{$C_{\mathrm{\!Q\!D\!V}}$} are normalization parameters defined by \scalebox{0.8}{${{\left[{I\!_0}\!\left({{E'}\!_J}\right)\!{I\!_0}\!\left( {E'}\!_{J}\!^0 \right) \right]}\!^{{M_x}\!{M_{\tau }}}}$} and \scalebox{0.8}{${{\left[{I\!_0}\!\left({{E'}\!_S}\! \right)\!{{I}\!_{0}}\!\left({E'}\!_{S}\!^0\right) \right]}\!^{{M_x}\!{M_{\tau }}}}$}, respectively. Both \scalebox{0.9}{$b_{i}(x,\tau)$} and \scalebox{0.9}{$\widetilde{b}_{i}(x,\tau)$} are auxiliary magnetic fields with integer values. Dual integer value fields \scalebox{0.9}{$f_{j}(x,\tau)$} and \scalebox{0.9}{$\widetilde{f}_{j}(x,\tau)$} are introduced to \scalebox{0.9}{$b_{i}(x,\tau)$} and \scalebox{0.9}{$\widetilde{b}_{i}(x,\tau)$}, respectively, as follows \cite{ref30}:
\begin{align}\label{eq62}
\scalebox{0.95}{$\displaystyle
b_i(x,\tau)\equiv\epsilon_{ij}f_{j}(x,\tau)
$},
\end{align}
\begin{align}\label{eq63}
\scalebox{0.95}{$\displaystyle
\widetilde{b}_i(x,\tau)\equiv\epsilon_{ij}\widetilde{f}_{j}(x,\tau)
$},
\end{align}
where \scalebox{0.9}{${{\varepsilon }_{0x}}\!\!=\!\!-{{\varepsilon }_{x0}}\!=\!1$} is the Levi–Civita symbol of two dimensions. 
By using the dual transformations of Eq.(\ref{eq62}) and (\ref{eq63}), the following equations are obtained for Eq.(\ref{eq60}) and (\ref{eq61}): 
\begin{align}\label{eq64}
\scalebox{0.83}{$\displaystyle
Z_{\!Q\!V}\!\!=\!\!{C\!_{\!Q\!V}}\sum_{\left\{{f}\right\}}{{\delta}_{{\nabla}_j\epsilon_{jk}f_{k},0}}\exp\sum_{x,\tau}\!
\left[
\frac{-f_{0}^{2}(x,\tau)}{2{{\left({E'}\!_{J}^{0}\right)}_v}}+\frac{-f_{x}^{2}(x,\tau)}{2{{\left({{E'}\!\!_J} \right)}_v}}
\right]
$},
\end{align}
\begin{align}\label{eq65}
\scalebox{0.83}{$\displaystyle
Z_{\!Q\!D\!V}\!\!=\!\!{C\!_{\!Q\!D\!V}}\!\!\!\sum_{\left\{\tilde{f}\right\}}\!{{\delta}_{{\nabla}_j\epsilon_{jk}{\tilde{f}_k},0}}\!\exp\!\sum_{x,\tau}\!
\left[
\!\frac{-\tilde{f}_{0}^{2}(x,\tau)}{2{{\left({E'}\!_{S}^{0}\right)}_v}}+\frac{-\tilde{f}_{x}^{2}(x,\tau)}{2{{\left({{E'}\!\!_S} \right)}_v}}
\right]
$},
\end{align}
Introducing the Poisson's formula of Eq.(\ref{eq66}) into  (\ref{eq64}) and  (\ref{eq65}), yields Eq.(\ref{eq67}) and (\ref{eq68}), respectively. 
\begin{align}\label{eq66}
\scalebox{0.72}{$\displaystyle
\sum_{\left({f_j}\right)=-\infty}^{\infty}\!\!\!\!\!\!{{\delta_{{\nabla _j}{{\varepsilon}_{jl}}{f_l},0}}\left(\cdot\cdot\right)}\!=\!\!\!\int\limits_{-\infty }^{\infty}\!\!\!d{B_j}\!\left(\cdot\cdot\right)\!\!\!\!\!\!\sum_{\left({l_j}\right)=-\infty}^{\infty}\!\!\!\!\!\!{{{\delta }_{{{\nabla }_j}{{\varepsilon }_{jl}}{l_l},0}}\exp\!\sum_{x,\tau}\!\left(\!i2\pi\!\sum\limits_{j=0}^{2}{{{\varepsilon }_{jl}}{l_l}{B_j}}\!\right)}
$},
\end{align}
\begin{align}\label{eq67}
\scalebox{0.80}{$\displaystyle
Z_{QV}\!\!\equiv\!\!C_{\mathrm{QV}}\!\sum_{\left\{l\right\}}\delta_{\nabla_j\epsilon_{jk}l_k,0}\!\!\int\!\!\!D\!B_0\!\!\int\!\!\!D\!B_x\exp\!\sum_{x,\tau}\!
\left[\frac{-B^2_0(x,\!\tau)}{2\left(E^{\prime}_{J}\right)_v}\!
\right.
$}\nonumber\\
\scalebox{0.80}{$\displaystyle
\left.
+\!\frac{-B^2_x(x,\!\tau)}{2\left(E^{\prime 0}_{J}\right)_v}
\!+\!2\pi l_x(x,\tau)B_0(x,\tau)\!+\!i 2\pi l_0(x,\tau)B_x(x,\tau)
\right]
$},
\end{align}
\begin{align}\label{eq68}
\scalebox{0.80}{$\displaystyle
Z_{QDV}\!\!\equiv\!\!C_{\mathrm{QDV}}\!\sum_{\left\{\tilde{l}\right\}}\delta_{\nabla_j\epsilon_{jk}\tilde{l}_k,0}\!\!\int\!\!\!D\!\tilde{B}_0\!\!\int\!\!\!D\!\tilde{B}_x\exp\!\sum_{x,\tau}\!
\left[\frac{-\tilde{B}^2_0(x,\!\tau)}{2\left(E^{\prime}_{S}\right)_v}\!
\right.
$}\nonumber\\
\scalebox{0.80}{$\displaystyle
\left.
+\!\frac{-\tilde{B}^2_x(x,\!\tau)}{2\left(E^{\prime 0}_{S}\right)_v}
\!+\!2\pi \tilde{l}_x(x,\tau)\tilde{B}_0(x,\tau)\!+\!i 2\pi \tilde{l}_0(x,\tau)\tilde{B}_x(x,\tau)
\right]$},
\end{align}
Integrating over the continuous value fields ${{B}_{j}}$and ${{\tilde{B}}_{j}}$ of Eq.(\ref{eq67}) and (\ref{eq68}) yields the following equations:
\begin{align}\label{eq69}
\scalebox{0.80}{$\displaystyle
Z_{QV}\!\!=\!\!R_{QV}\!\!\sum_{\left\{l\right\}}\!{{\delta_{{\nabla_j}{\varepsilon _{jl}}{l_l},0}}}\exp\!\!\sum_{x,\tau}\!\Bigl[
\!-2{{\pi\!}^2}{{\left({E'}\!_J^0 \right)\!}_v}{l_0\!}^2\!-\!2{{\pi\!}^2}{{\left(E'\!_J\right)\!}_v}{l_x\!}^2
\Bigr]
$},
\end{align}
\begin{align}\label{eq70}
\scalebox{0.80}{$\displaystyle
Z_{\!Q\!D\!V}\!\!=\!\!R_{\!Q\!D\!V}\!\!\sum_{\left\{\tilde{l}\right\}}\!{{\delta_{{\nabla_j}{\varepsilon _{jl}}{\tilde{l}_l},0}}}\exp\!\!\sum_{x,\tau}\!\!\left[
\!-2{{\pi\!}^2}{{\left({E'}\!_S^0 \right)\!}_v}{\tilde{l}_0\!}^2\!\!-\!2{{\pi\!}^2}{{\left(E'\!_S\right)\!}_v}{\tilde{l}_x\!}^2
\right]
$},
\end{align}
The Kronecker deltas, when rewritten in the integral form, allow the equations to be written as follows:
\begin{align}\label{eq71}
\scalebox{0.78}{$\displaystyle
Z_{QV}\!\!=\!\!R_{QV}\!\!\sum_{\left\{l\right\}}\!\!\int\!\!D\tilde{\theta}\exp\!\!\sum_{x,\tau}\!\Bigl[
\!-\!2{{\pi\!}^2}{{\left({E'}\!_J^0 \right)\!}_v}{l_0\!}^2\!-\!2{{\pi\!}^2}{{\left(E'\!_J\right)\!}_v}{l_x\!}^2
\!\!-\!\!i{{\nabla }_j}{{\varepsilon }_{jl}}l\tilde{\theta }
\Bigr]
 $},
\end{align}
\begin{align}\label{eq72}
\scalebox{0.78}{$\displaystyle
Z_{\!Q\!D\!V}\!\!=\!\!R_{\!Q\!D\!V}\!\!\sum_{\left\{\tilde{l}\right\}}\!\!\!\int\!\!\!D\theta\exp\!\!\sum_{x,\tau}\!\!\left[
\!-\!2{{\pi\!}^2}{\!{\left({E'}\!_S^0 \right)\!}_v}{\tilde{l}_0\!}^2\!-\!2{{\pi\!}^2}{\!{\left(E'\!_S\right)\!}_v}{\tilde{l}_x\!}^2
\!\!-\!\!i{{\nabla }_j}{{\varepsilon }_{jl}}\tilde{l}\theta
\right]
 $},
\end{align}
Using the identity of Eq.(\ref{eq59}) for (\ref{eq71}) and (\ref{eq72}), respectively, the equations become: 
\begin{align}\label{eq73}
\scalebox{0.80}{$\displaystyle
Z_{QV}\!\!=\!\!C_{QV}\!\!\sum_{\left\{\tilde{n}\right\}}\!\!\int\!\!\!D\tilde{\theta}\exp\!\!\sum_{x,\tau}\!\left[
\frac{\!-\!\Bigl(\!{\nabla }_{\tau}\tilde{\theta}\!-\!2\pi {\tilde{n}_0}\!\Bigr)\!^2}{8{\pi\!}^2{{\left({{E'}\!\!_J} \right)}_v}}
\!+\!\frac{\!-\!\Bigl(\!{\nabla }_{x}\tilde{\theta}\!-\!2\pi {\tilde{n}_x}\!\Bigr)\!^2}{8{\pi\!}^2{{\left({E'}\!_{J}^{0}\right)}_v}}
\right]
 $},
\end{align}
\begin{align}\label{eq74}
\scalebox{0.80}{$\displaystyle
Z_{\!Q\!D\!V}\!\!=\!\!C_{\!Q\!D\!V}\!\!\sum_{\left\{{n}\right\}}\!\!\int\!\!\!D\theta\exp\!\!\sum_{x,\tau}\!\!\left[
\frac{\!-\!\Bigl(\!{\nabla }_{\tau }\theta\!-\!2\pi {n_0}\!\Bigr)\!^2}{8{\pi\!}^2{{\left({{E'}\!\!_S} \right)}_v}}\!+\!\frac{\!-\!\Bigl(\!{\nabla }_{x}\theta\!-\!2\pi {n_x}\!\Bigr)\!^2}{8{\pi\!}^2{{\left({E'}\!_{S}^{0}\right)}_v}}
\right]
 $},
\end{align}
By using the inverse transform of the Villain approximation introduced in Eq.(\ref{eq57}) and (\ref{eq58}) on Eq.(\ref{eq73}) and (\ref{eq74}), respectively, the equations can be rewritten as follows: 
\begin{align}\label{eq75}
\scalebox{0.80}{$\displaystyle
Z_{QV}\!\!\approx\!\!{C'_{QV}}\!\!\int\!\!\!D\tilde{\theta}\exp\!\!\sum_{x,\tau}\!\left(\!
\frac{1}{4{{\pi }^2}{{E'}_J}}\cos\!{{\nabla}_{\tau}}\tilde{\theta}+\frac{1}{4{{\pi }^2}{E'}_J^0}\cos\!{{\nabla }_x}\tilde{\theta}
\right)
 $},
\end{align}
\begin{align}\label{eq76}
\scalebox{0.80}{$\displaystyle
Z_{\!Q\!D\!V}\!\!\approx\!\!{C'_{\!Q\!D\!V}}\!\!\int\!\!\!D{\theta}\exp\!\!\sum_{x,\tau}\!\left(\!
\frac{1}{4{{\pi }^2}{{E'}_S}}\cos\!{{\nabla}_{\tau}}{\theta}+\frac{1}{4{{\pi }^2}{E'}_S^0}\cos\!{{\nabla }_x}{\theta}
\right)
 $},
\end{align}
where \scalebox{0.78}{$C'_{QV}$} and \scalebox{0.78}{$C'_{\!Q\!D\!V}$} are defined by \scalebox{0.78}{${C_{\!Q\!V}}\!/\!{{R_v}\!(\!1\!/4{{\pi }^2}\!{E'}\!_J^0){R_v}( \!1\!/4{{\pi}^2}\!{E'}\!\!_J)}$} and \scalebox{0.8}{${C_{\!Q\!D\!V}}\!/\!{{R_v}\!(\!1\!/4{{\pi }^2}\!{E'}\!_S^0){R_v}(\!1\!/4{{\pi}^2}\!{E'}\!\!_S)}$}, respectively. The following equation is derived from Eq.(\ref{eq37}), (\ref{eq38}), and (\ref{eq16}):  
\begin{align}\label{eq77}
\scalebox{0.90}{$\displaystyle
{E'}_{J}^{0}\equiv\frac{1}{4{{\pi }^2}{E'_S}}
 $},
\end{align}
\begin{align}\label{eq78}
\scalebox{0.90}{$\displaystyle
{E'}_{S}^{0}\equiv\frac{1}{4{{\pi }^2}{E'_J}}
 $},
\end{align}
When the relationships of Eq.(\ref{eq77}) and (\ref{eq78}) are used in Eq.(\ref{eq75}) and (\ref{eq76}), the following equation can be derived:
\begin{align}\label{eq79}
\scalebox{0.76}{$\displaystyle
Z'\!\!_{A\!X\!Y}\!\!\approx\!\!Z_{Q\!V}\!\!\approx\!\!\frac{C\!_{Q\!V}}{R\!_{Q\!D\!V}}\!\!\!\int\!\!\!D\tilde{\theta}\exp\!\!\sum_{x,\tau}\!\!\left(\!
{E'}\!_{S}^{0}\!\cos\!{{\nabla}\!_{\tau}}\tilde{\theta}\!+\!E'\!\!_S\!\cos\!{{\nabla}\!_x}\tilde{\theta}
\right)\!\!=\!\!\frac{C\!_{Q\!V}}{R\!_{Q\!D\!V}}\!Z'\!\!_{D\!A\!X\!Y}
 $},
\end{align}
\begin{align}\label{eq80}
\scalebox{0.76}{$\displaystyle
Z'\!\!_{D\!A\!X\!Y}\!\!\approx\!\!Z_{Q\!P\!S\!V}\!\!\approx\!\!\frac{C\!_{Q\!D\!V}}{R\!_{D\!V}}\!\!\!\int\!\!\!D{\theta}\exp\!\!\sum_{x,\tau}\!\!\left(\!
{E'}\!_{J}^{0}\!\cos\!{{\nabla}\!_{\tau}}{\theta}\!+\!E'\!\!_J\!\cos\!{{\nabla}\!_x}{\theta}\right)\!\!=\!\!\frac{C\!_{Q\!D\!V}}{R\!_{Q\!V}}\!Z'\!\!_{A\!X\!Y}
 $},
\end{align}
In Eq.(\ref{eq79}) and (\ref{eq80}), it is guaranteed that  ${Z'_{\!A\!X\!Y}} $ and  ${Z'_{\!D\!A\!X\!Y}} $ are completely dual relationships under the following condition regarding normalization parameters:
\begin{align}\label{eq81}
\scalebox{0.76}{$\displaystyle
 \frac{{{C}_{QV}}}{{{R}_{QDV}}}\frac{{{C}_{QDV}}}{{{R}_{QV}}}\approx 1                         
 $}.
\end{align}
$ $
\section{Ginzburg-Landau theory and Kosterlitz-Thouless transition \label{sec8}}
In this section, starting from the two partition functions ${Z'_{\!A\!X\!Y}}$ and ${Z'_{\!D\!A\!X\!Y}}$of $J\!J$ and $Q\!P\!S$ from Eq.(\ref{eq55}) and (\ref{eq56}) which are dual to each other, we consider the Ginzburg-Landau theory ($GL$ theory) of  two types and the Kosterlitz-Thouless transition ($KT$ transition) of two types. For each of Eq.(\ref{eq55}) and (\ref{eq56}), we introduce two element unit vectors \scalebox{0.8}{${{U}_{l}}\!=\!\!\Bigl[ \cos \theta,\sin \theta \Bigr]$} and \scalebox{0.8}{${{\tilde{U}}_{l}}\!=\!\!\left[ \cos \tilde{\theta },\sin \tilde{\theta }\right]$}\scalebox{0.9}{$(l\! = \!1,2)$} respectively as follows:
\begin{align}\label{eq82}
\scalebox{0.76}{$\displaystyle
Z'_{\!A\!X\!Y\!}\left( {E'}_J^0, {E'}_J  \right)\!\!=\!\!\int\!\!\!D{\theta}\exp\!\left\{ {{E'}\!\!_J}d\sum\limits_{x}{\sum\limits_{l=1}^{2}{{U_l}\left( x,\tau  \right)R{U_l}\left( x,\tau  \right)}} \right\}
 $},
\end{align}
\begin{align}\label{eq83}
\scalebox{0.76}{$\displaystyle
Z'_{\!D\!A\!X\!Y\!}\left( {E'}_S^0, {E'}_S  \right)\!\!=\!\!\int\!\!\!D{\tilde{\theta}}\exp\!\left\{ {{E'}\!\!_S}\tilde{d}\sum\limits_{x}{\sum\limits_{l=1}^{2}{{\tilde{U}_l}\left( x,\tau  \right)\tilde{R}{\tilde{U}_l}\left( x,\tau  \right)}} \right\}
 $},
\end{align}
Where the lattice difference operators $R$ and  $\tilde{R}$ are respectively defined as\cite{ref30}:
\begin{align}\label{eq84}
\scalebox{0.76}{$\displaystyle
R\equiv\!1\!+\!\frac{1}{2d}\!\left( {{{\bar{\nabla }}}_{x}}{{\nabla }_{x}}+\gamma {{{\bar{\nabla }}}_{\tau }}{{\nabla }_{\tau }} \right), \gamma \equiv \frac{{E'}\!_{J}^0}{{E'}\!\!_J}
$},
\end{align}
\begin{align}\label{eq85}
\scalebox{0.76}{$\displaystyle
\tilde{R}\equiv\!1\!+\!\frac{1}{2\tilde{d}}\!\left( {{{\bar{\nabla }}}_{x}}{{\nabla }_{x}}+\tilde{\gamma} {{{\bar{\nabla }}}_{\tau }}{{\nabla }_{\tau }} \right), \tilde{\gamma} \equiv \frac{{E'}\!_{S}^0}{{E'}\!\!_S}
$},
\end{align}
Were $d\!\!\equiv\!\!1+\!\gamma $ and $\tilde{d}\!\!\equiv\!\!1+\!\tilde{\gamma }$ are anisotropic dimensional constants of $J\!J$ and $Q\!P\!S$, respectively.
Moreover in Eq.(\ref{eq82}), we introduce two sets of real two component fields ${u_l}$ and ${{\psi }_l}$ \scalebox{0.9}{$(l\! = \!1,2)$}  which satisfy the following identity\cite{ref30}:
\begin{align}\label{eq86}
\scalebox{0.76}{$\displaystyle
\int\limits_{-\infty }^{\infty }\!\!\!{d{u_1}d{u_2}}\!\!\!\!\int\limits_{-i\infty }^{i\infty }\!\!{\frac{d{{\psi }_1}d{{\psi }_2}}{{{\left( 2\pi i \right)}^2}}\exp\!\left\{ -{{\psi }_l}\left( {u_l}-{U_l} \right) \right\}}\!=\!\!\!\int\limits_{-\infty }^{\infty }\!\!\!{d{u_1}d{u_2}{{\delta }^2}\!\left( {u_l}-{U_l} \right)}\!=\!1,
$}\end{align}
\begin{align}\label{eq87}
\scalebox{0.88}{$\displaystyle
{{Z'}\!_{\!A\!X\!Y}}\!\!=\!\!\prod\limits_{x,\tau ,l}\!\!{\int\limits_{-\infty }^{\infty }\!\!{d{{u}_{l}}}}\!\!\!\int\limits_{-i\infty }^{i\infty }\!\!{\frac{d{{\psi }_l}}{{{\left( 2\pi i \right)}^{2}}}}\exp\!\!\sum\limits_{x,\tau ,l}\!{\left\{{{E'}\!\!_J}d{u_l}\left( x,\tau  \right)R{u_l}\left(x,\tau \right)\right.}
$}\nonumber\\
\scalebox{0.88}{$\displaystyle
\left.-{{\psi }_{l}}\left( x,\tau  \right){{u}_{l}}\left( x,\tau  \right)+\ln {{I}_{0}}\left( \left| {{\psi }_{l}}\left( x,\tau  \right) \right| \right) \right\}
$},\quad\quad\quad\quad\end{align}
where we have used that the product of $\theta$ functional integrals is given as follows:
\begin{align}\label{eq88}
\scalebox{0.80}{$\displaystyle
\prod\limits_{x,\tau }\!\!{ \int\limits_{-\pi }^{\pi }\!\!{\frac{d\theta}{2\pi }}}\!\exp\!\!\left[ \!\sum\limits_{x,\tau ,l}{{{\psi }_l}\left( x,\tau  \right){U_l}\left( x,\tau  \right)}\! \right]\!\!=\!\exp\!\!\sum\limits_{x,\tau ,l}\!\!{\left\{ \ln {I_0}\left( \left| {{\psi }_l}\left( x,\tau  \right) \right| \right) \right\}}
$},
\end{align}
where  ${I_0}\!\left( \left| \psi  \right| \right)$ ($\left| \psi  \right|\!\equiv\!\sqrt{{{\psi }_1}\!^2+{{\psi }_2}\!^2}$ ) is the modified Bessel functions of integer 0th order. In Eq.(\ref{eq87}), performing the integrals over ${u_l}$ fields, we obtain the partition function by the complex field $\psi\!\equiv\!{{\psi }_1}\!+\!i{{\psi }_2}$ and ${{\psi }^{*}}\!\equiv\!{{\psi }_1}\!-\!i{{\psi }_2}$.
\begin{align}\label{eq89}
\scalebox{0.80}{$\displaystyle
{{{Z'}}\!_{A\!X\!Y}}\!\!=\!\prod\limits_{x,\tau }{\left[ \int\limits_{-\pi }^{\pi }{\frac{d\psi \left( x,\tau  \right)d{{\psi }^{*}}\left( x,\tau  \right)}{4\pi {{{E'}}\!_J}d}} \right]}\exp \left\{ -{F'}\left( \psi ,{{\psi }^{*}} \right) \right\}
$},
\end{align}
\begin{align}\label{eq90}
\scalebox{0.80}{$\displaystyle
{F'}\left( \psi ,{{\psi }^{*}} \right)=\sum\limits_{x,\tau }{\left\{ \frac{1}{4{{E'}\!_J}d}{{\left| \psi \left( x,\tau  \right) \right|}^{2}}-\ln {{I}_{0}}\left[ \left| \hat{\psi }\left( x,\tau  \right) \right| \right] \right\}}
$},
\end{align}
\begin{align}\label{eq91}
\scalebox{0.80}{$\displaystyle
\hat{\psi }\left( x,\tau  \right)\!\equiv\! {{\hat{R}}^{\frac{1}{2}}}\psi \left( x,\tau  \right)\!=\!\sqrt{1\!+\!\frac{1}{2d}\!\left( {{{\bar{\nabla }}}_{x}}{{\nabla }_{x}}\!+\!\gamma {{{\bar{\nabla }}}_{\tau }}{{\nabla }_{\tau }} \right)}\psi \left( x,\tau  \right)
$},
\end{align}
In Eq.(\ref{eq90}), since $\psi$ and ${{\psi }^{*}}$ can be regarded as the order parameter of superconductivity\cite{ref30}, the dimensionless energy ${F}'\left( \psi ,{{\psi }^{*}} \right)$ can be Landau expansion of terms up to ${{\left| \psi  \right|}^{4}}$ and ${{\left| {{\nabla }_{x}}\psi  \right|}^{2}}$ as follouws\cite{ref30}:
\begin{align}\label{eq92}
\scalebox{0.76}{$\displaystyle
{{{F'}}\!_{GL}}\!\left( \psi ,{{\psi }^{*}} \right)\!=\!\!\sum\limits_{x,\tau }\!{\left\{ \frac{1}{8d}\!\left[ {{\left|{{\nabla }_{x}}\psi\right|}^{2}}\!\!+\!\gamma {{\left| {{\nabla }_{\tau }}\psi  \right|}^{2}} \right]\!\!+\!\frac{1}{4}\!\left( \!\frac{1}{{{E'}\!_J}d}\!-\!1\! \right)\!{{\left| \psi  \right|}^{2}}\!+\!\frac{1}{64}{{\left| \psi  \right|}^{4}} \right\}}
$},
\end{align}
${{F'}\!_{G\!L}}$ is Ginzburg-Landau ($GL$) energy of superconductivity or Pitaevskii energy of Superfluid in $1\!+\!\gamma $ dimension at zero temperature. 
Similarly, when $GL$ energy is calculated from Eq.(\ref{eq83}), it becomes as follows:
\begin{align}\label{eq93}
\scalebox{0.76}{$\displaystyle
{{\tilde{F'}}\!_{D\!G\!L}}\!\left( \tilde{\psi },{{{\tilde{\psi }}}^{*}} \right)\!\!=\!\!\sum\limits_{x,\tau }\!{\left\{ \frac{1}{8\tilde{d}}\!\!\left[ {{\left| {{\nabla }_{x}}\tilde{\psi } \right|}^{2}}\!\!+\!\!\tilde{\gamma }{{\left| {{\nabla }_{\tau }}\tilde{\psi } \right|}^{2}} \right]\!\!+\!\frac{1}{4}\!\left(\!\frac{1}{{{E'}\!_S}\tilde{d}}\!-1\!\right)\!{{\left| {\tilde{\psi }} \right|}^{2}}\!+\!\frac{1}{64}{{\left| {\tilde{\psi }} \right|}^{4}} \right\}}
$},
\end{align}
${{\tilde{F'}}\!_{\!D\!G\!L}}$ is Dual Ginzburg-Landau ($D\!G\!L$) energy of superinsulator $1\!+\!\tilde{\gamma }$ dimension at zero temperature. Here $\tilde{\psi }$ and ${{\tilde{\psi }}^{*}}$ can be regarded as order parameters of superinsulator. As opposed to being a condensate of $2e$ in which the order parameter of the superconductor is twice the elementary charge e, the order parameter of the superinsulator can be thought of as a condensate of $2{{\phi }_0}$ which is twice the quantum vortex ${{\phi }_0}\!\!=\!h/{\left( 2e \right)}$. From Eq.(\ref{eq92}) and (\ref{eq93}), the critical values ${{E'}\!_J\!}^{\!G\!L}$ and ${{E'}\!_S\!}^{D\!G\!L}$ by mean field approximation of ${{E'}\!\!_J}$ and ${{E'}\!\!_S}$ are as follows:
\begin{align}\label{eq94}
\scalebox{0.9}{$\displaystyle
{{E'}\!_J}^{\!G\!L}\!=\!{{d}^{-1}}\!=\!\frac{1}{1+\gamma }
$},
\end{align}
\begin{align}\label{eq95}
\scalebox{0.9}{$\displaystyle
{{E'}\!_S}^{D\!G\!L}\!=\!{{\tilde{d}}^{-1}}\!=\!\frac{1}{1+\tilde{\gamma }}  
$},
\end{align}
On the other hand, $A\!X\!Y$ model of $1\!+\!\gamma$ dimension and $D\!A\!X\!Y$ model of $1\!+\!\tilde{\gamma }$ dimension becomes a pseudo two-dimensional $X\!Y$ model under the condition of ${E'}\!_J^0\!=\!{{E'}\!_J}\left( \gamma \!=\!1 \right)$ and ${E'}\!_S^0\!=\!{{E'}\!_S}\left( \tilde{\gamma }\!=\!1 \right)$ respectively.Therefore, it is possible for the $A\!X\!Y$ model and the $D\!A\!X\!Y$ model to generate KT transition in the pseudo two dimension at zero temperature. In this case, the critical values of quantum $K\!T\! (Q\!K\!T)$ transition and dual quantum $K\!T\! (D\!Q\!K\!T)$ transition are as follows respectively:
\begin{align}\label{eq96}
\scalebox{0.9}{$\displaystyle
{{E'}\!_J}^{\!Q\!K\!T}\!=\!\frac{\text{2}}{\pi }
$},
\end{align}
\begin{align}\label{eq97}
\scalebox{0.9}{$\displaystyle
{{E'}\!_S}^{\!D\!Q\!K\!T}=\!\frac{\text{2}}{\pi }
$},
\end{align}
 Where, ${{E'}\!_J}^{\!Q\!K\!T}$ represents the critical value of ${{E'}\!_J}$ due to the $Q\!K\!T$ transition. $Q\!K\!T$ transition is a topological phase transition due to the vortex condensation in pseudo two dimensional space. On the other hand, ${{E'}\!_S}^{\!Q\!D\!K\!T}$ represents the critical value of ${{E'}\!_S}$ due to the $D\!Q\!K\!T$ transition. TABLE I shows the critical values of ${{E'}\!_J}$, ${{E'}\!_L}$ and ${{E'}\!_S}$ according to $Q\!K\!T$ and $D\!Q\!K\!T$ transition. 
\begin{table}[htbp]
\caption{\label{tab:table1}%
Critical value of ${{E'}\!_J}$, ${{E'}\!_L}$ and ${{E'}\!_S}$ by the transition of $Q\!K\!T$ and $D\!Q\!K\!T$.
}
\begin{ruledtabular}
\begin{tabular}{ccc}
\quad\quad\textrm{QKT}&
\textrm{DQKT}&\\
\colrule
\quad\quad\quad\quad${{E'}\!\!_J}^{\!Q\!K\!T}\!\!={\text{2}}/{\pi }$&${{E'}\!_S}^{\!D\!Q\!K\!T}\!\!={\text{2}}/{\pi }$\\
\quad\quad\quad\quad${{E'}\!\!_L}^{\!Q\!K\!T}\!\!=\!{\text{4}}{\pi }$ & ${{E'}\!_c}^{\!D\!Q\!K\!T}={\text{4}}{\pi }$\\
\quad\quad\quad\quad${{E'}\!\!_S}^{\!Q\!K\!T}\!\!=\!{\text{1}}/{8\pi }$ & ${{E'}\!_J}^{\!D\!Q\!K\!T}={\text{1}}/{8\pi }$\\
\end{tabular}
\end{ruledtabular}
\end{table}
 
 Similarly, from Eq.(\ref{eq94}) and Eq.(\ref{eq95}), the critical values of the mean field approximation under the conditions of $\gamma\!\!=\!\!1$ and $\tilde{\gamma}\!\!=\!\!1$ are respectively as follows:
\begin{align}\label{eq98}
\scalebox{0.9}{$\displaystyle
{{E'}\!_J}^{\!G\!L}\!\left( \gamma \!=\!1 \right)=\frac{\text{1}}{\text{2} }
$},
\end{align}
\begin{align}\label{eq99}
\scalebox{0.9}{$\displaystyle
{{E'}\!_S}^{\!D\!G\!L}\!\left( \tilde{\gamma }\!=\!1 \right)=\frac{\text{1}}{\text{2} }
$},
\end{align}
\begin{table}[htbp]
\caption{\label{tab:table2}%
Critical value of ${{E'}\!_J}$, ${{E'}\!_L}$ and ${{E'}\!_S}$ by the transition of $G\!L$ theory and $D\!G\!L$ theory under the conditions of $\gamma\!\!=\!\!1$ and $\tilde{\gamma}\!\!=\!\!1$.
}
\begin{ruledtabular}
\begin{tabular}{ccc}
\quad\quad\textrm{GL}&
\textrm{DGL}&\\
\colrule
\quad\quad\quad\quad${{E'}\!\!_J}^{\!G\!L}\!\!=\text{1}/\text{2} $&${{E'}\!_S}^{\!D\!G\!L}\!\!=\text{1}/\text{2} $\\
\quad\quad\quad\quad${{E'}\!\!_L}^{\!G\!L}\!\!=\!{\pi }^2$ & ${{E'}\!_c}^{\!D\!G\!L}={\pi }^2$\\
\quad\quad\quad\quad${{E'}\!\!_S}^{\!G\!L}\!\!=\!{\text{1}}/2{\pi }^2$ & ${{E'}\!_J}^{\!D\!G\!L}={\text{1}}/2{\pi }^2$\\
\end{tabular}
\end{ruledtabular}
\end{table}
TABLE I\hspace{-.1em}I shows the critical values of ${{E'}\!_S}$, ${{E'}\!_c}$ and ${{E'}\!_J}$ according to $G\!L$ theory and $D\!G\!L$ theory under the conditions of $\gamma\!\!=\!\!1$ and $\tilde{\gamma}\!\!=\!\!1$. From Eq.(\ref{eq96}) and Eq.(\ref{eq98}), the difference between $Q\!K\!T$ transition and mean field approximation is about \scalebox{0.85}{${{E'}\!\!_J}^{\!Q\!K\!T}\!-\!{{E'}\!\!_J}^{\!G\!L}\!\!\approx\! 0.137$}. Similarly, the difference between Eq.(\ref{eq97}) and Eq.(\ref{eq99}) \scalebox{0.85}{${{E'}\!\!_S}^{\!D\!Q\!K\!T}\!\!-\!{{E'}\!\!_S}^{\!D\!G\!L}\!\!\approx\! 0.137$}. In the $2\!+\!1$ dimensional $J\!J$ array of  \cite{ref35}-\cite{ref37}, a self-dual model is obtained (in the sense of electromagnetic duality) by adding kinetic terms to the vortices. On the other hand, our $1\! +\!1$ dimensional model does not artificially add kinetic terms to the vortices. 
In other words, self-dual form mixed Chern-Simons action\cite{ref36}-\cite{ref37} is not assumed from the beginning. Instead, starting from the two Hamiltonian of $J\!J$ and $Q\!P\!S$ which are dual to each other, it is a method to calculate various physical quantities from them by assuming the existence of $A\!X\!Y$ model and $D\!A\!X\!Y$ model which are dual to each other. Since the $D\!G\!L$ energy of the superinsulator of Eq.(91) is considered to be the $G\!L$ energy of the vortex, it contains a kinematic vortex term in a different sense from Ref \cite{ref35}-\cite{ref37}.

\section{Summary and Conclusion \label{sec9}}
This section contains the summarized conclusions from each section in this paper as follows: 
Section 1: 
Two dual Hamiltonians were introduced into a quantum $L\!C$ circuit, known as the simplest quantum dual system, and the dual condition was applied between the current and the voltage of the electric circuit. Thus introducing a general theory and method for constructing a dual system named the $D\!H$ (dual Hamiltonian) method. 
Section 2:
The $D\!H$ method was applied between the $J\!J$ and the $Q\!P\!S\!J$ in a single junction, allowing the following to be derived: two relational expressions of particle number and phase between dual systems, $Q\!P\!S$ amplitude and charge per charge energy, and the relationship between Josephson energy and induced energy per flux quantum. Furthermore, kinetic inductance and kinematic capacitance were derived in a nonlinear form. This result is an extension of the result of Mooij et al\cite{ref11}-\cite{ref13}. 
Section 3: 
The relation between the Hamiltonian and the equivalent circuit of each quantum circuit was clarified by applying a simple proof of self-duality in various quantum junction circuits. 
Section 4:
Owing to the duality of the $J\!J$ and the $Q\!P\!S\!J$, the transition of superconductor–superconductor could be explained by simple consideration. This indicated the possibility that a $Q\!P\!S\!J$ could be constructed from the junction of two superinsulators. 
Section 5: 
The $J\!J$ and $Q\!P\!S\!J$ of a single junction were examined using the partition function and incorporating the quantum effect by path integration. By introducing the energy of the imaginary time component in the JJ and $Q\!P\!S\!J$, the duality between them was demonstrated and approximately established. 
Section 6: 
The partition function of \scalebox{0.9}{$A\!X\!Y$} model and \scalebox{0.9}{$D\!A\!X\!Y$} model was examined in $1\!+\!1$ dimensions equivalent to the $J\!J$ and $Q\!P\!S\!J$ in a one-dimensional nanowire. 
Section 7: 
Within the Villain approximation, it was confirmed that duality was accurately established. 
Overall results and conclusions: 
Section 8:  Starting with the two  partition functions of $J\!J$ and $Q\!P\!S$, which are dual each other, we have determined two critical values of two types of $G\!L$ theory and $K\!T$ transition, respectively. 
The most important result of this paper is that by introducing two Hamiltonians that were dual with each other, the $D\!H$ method was established, which is a general method for constructing an accurate dual system. The reliability of this method was proved accurately within the Villain approximation for the partition function of the $1\!+\!1$ dimensional \scalebox{0.9}{$A\!X\!Y$} model and \scalebox{0.9}{$D\!A\!X\!Y$} model corresponding to the $J\!J$ and $Q\!P\!S\!J$ in a one-dimensional nanowire system. It is believed that the $D\!H$ method will prove to be a very effective method for future research into the $Q\!P\!S$ and superinsulator phenomena.

\section{Acknowledgments\label{Acknowledgments}}
I would like to thank all the faculty and staff of Aichi University of Technology.


\begin{thebibliography}{37}
\expandafter\ifx\csname natexlab\endcsname\relax\def\natexlab#1{#1}\fi
\expandafter\ifx\csname bibnamefont\endcsname\relax
  \def\bibnamefont#1{#1}\fi
\expandafter\ifx\csname bibfnamefont\endcsname\relax
  \def\bibfnamefont#1{#1}\fi
\expandafter\ifx\csname citenamefont\endcsname\relax
  \def\citenamefont#1{#1}\fi
\expandafter\ifx\csname url\endcsname\relax
  \def\url#1{\texttt{#1}}\fi
\expandafter\ifx\csname urlprefix\endcsname\relax\def\urlprefix{URL }\fi
\providecommand{\bibinfo}[2]{#2}
\providecommand{\eprint}[2][]{\url{#2}}
\bibitem[{\citenamefont{Seiberg and Witten}(1994)}]{ref1}
\bibinfo{author}{\bibfnamefont{N.}~\bibnamefont{Seiberg}} \bibnamefont{and}
  \bibinfo{author}{\bibfnamefont{E.}~\bibnamefont{Witten}},
  \bibinfo{journal}{Nucl. Phys. B} \textbf{\bibinfo{volume}{426}},
  \bibinfo{pages}{19} (\bibinfo{year}{1994}).
\bibitem[{\citenamefont{Witten}(1988)}]{ref2}
\bibinfo{author}{\bibfnamefont{E.}~\bibnamefont{Witten}},
  \bibinfo{journal}{Commun. Math. Phys}
  \textbf{\bibinfo{volume}{117}},
 \bibinfo{pages}{353} (\bibinfo{year}{1988}).
\bibitem[{\citenamefont{Witten}(1997)}]{ref3}
\bibinfo{author}{\bibfnamefont{E.}~\bibnamefont{Witten}},
  \bibinfo{journal}{Phys. Today 50}
  \textbf{\bibinfo{volume}{5}},
 \bibinfo{pages}{28} (\bibinfo{year}{1997}).
\bibitem[{\citenamefont{Kogut}(1979)}]{ref4}
\bibinfo{author}{\bibfnamefont{J.~B.} \bibnamefont{Kogut}}, 
\bibinfo{journal}{Rev. Mod. Phys} \textbf{\bibinfo{volume}{51}},
  \bibinfo{pages}{659} (\bibinfo{year}{1979}).
\bibitem[{\citenamefont{Savit}(1980)}]{ref5}
\bibinfo{author}{\bibfnamefont{R.}~\bibnamefont{Savit}}, 
\bibinfo{journal}{Rev. Mod. Phys} \textbf{\bibinfo{volume}{52}},
  \bibinfo{pages}{453} (\bibinfo{year}{1980}).
\bibitem[{\citenamefont{Kramers and Wannier}(1941{\natexlab{a}})}]{ref6}
\bibinfo{author}{\bibfnamefont{H.~A.} \bibnamefont{Kramers}}
  \bibnamefont{and} \bibinfo{author}{\bibfnamefont{G.~H.}
  \bibnamefont{Wannier}}, \bibinfo{journal}{Phys. Rev}
  \textbf{\bibinfo{volume}{60}}, \bibinfo{pages}{252}
  (\bibinfo{year}{1941}{\natexlab{a}}).
\\ \bibinfo{author}{\bibfnamefont{H.~A.} \bibnamefont{Kramers}}
\bibnamefont{and} \bibinfo{author}{\bibfnamefont{G.~H.}
\bibnamefont{Wannier}}, \bibinfo{journal}{Phys. Rev}
\textbf{\bibinfo{volume}{60}}, \bibinfo{pages}{263}
 (\bibinfo{year}{1941}{\natexlab{a}}).
\bibitem[{\citenamefont{Fisher}(2004)}]{ref7}
\bibinfo{author}{\bibfnamefont{M.~P.~A}~\bibnamefont{Fisher}},
  \emph{\bibinfo{title}{Duality in low dimensional quantum field theories, Strong interactions in low dimensions}},
\bibinfo{pages}{419-438}(\bibinfo{publisher}{Springer}, \bibinfo{year}{2004}).
\bibitem[{\citenamefont{Cherry}(1949)}]{ref8}
\bibinfo{author}{\bibfnamefont{R.}~\bibnamefont{Cherry}}, 
\bibinfo{journal}{Proc. Physical Societ} \textbf{\bibinfo{volume}{62B}},
  \bibinfo{pages}{101-111} (\bibinfo{year}{1949}).
\bibitem[{\citenamefont{Leon et~al.}(2012)\citenamefont{Leon, Farazmand, and Joseph}}]{ref9}
\bibinfo{author}{\bibfnamefont{F.}~\bibnamefont{Leon}},
  \bibinfo{author}{\bibfnamefont{A.}~\bibnamefont{Farazmand}}, \bibnamefont{and}
  \bibinfo{author}{\bibfnamefont{P.}~\bibnamefont{Joseph}},
  \bibinfo{journal}{IEEE Transactions on Power Delivery} \textbf{\bibinfo{volume}{27(4)}},
  \bibinfo{pages}{2390-2398} (\bibinfo{year}{2012}).
\bibitem[{\citenamefont{Marino et~al.}(2012)\citenamefont{Marino, Leon, and Fernandez}}]{ref10}
\bibinfo{author}{\bibfnamefont{C.}~\bibnamefont{Alverez-Marino}},
  \bibinfo{author}{\bibfnamefont{F. de.}~\bibnamefont{Leon}}, \bibnamefont{and}
  \bibinfo{author}{\bibfnamefont{X. M.}~\bibnamefont{Lopez-Fernandez}},
  \bibinfo{journal}{IEEE Transactions on Power Delivery} \textbf{\bibinfo{volume}{27(1)}},
  \bibinfo{pages}{353-361} (\bibinfo{year}{2012}).
\bibitem[{\citenamefont{Mooij and Harmans}(2005)}]{ref11}
\bibinfo{author}{\bibfnamefont{J.E.}~\bibnamefont{Mooij}} \bibnamefont{and}
  \bibinfo{author}{\bibfnamefont{C.J.P.M.}~\bibnamefont{Harmans}},
  \bibinfo{journal}{New Journal of Physics} \textbf{\bibinfo{volume}{7}},
  \bibinfo{pages}{219} (\bibinfo{year}{2005}).
\bibitem[{\citenamefont{Mooij and Nazarov}(2006)}]{ref12}
\bibinfo{author}{\bibfnamefont{J.E.}~\bibnamefont{Mooij}} \bibnamefont{and}
  \bibinfo{author}{\bibfnamefont{Yu. V.}~\bibnamefont{Nazarov}},
  \bibinfo{journal}{Nature Physics} \textbf{\bibinfo{volume}{2}},
  \bibinfo{pages}{169-172} (\bibinfo{year}{2006}).
\bibitem[{\citenamefont{Mooij J. E. et al}(2015)}]{ref13}
\bibinfo{author}{\bibfnamefont{J.E.}~\bibnamefont{Mooij, et al.,}} \bibnamefont{}
  \bibinfo{journal}{New J. Phys} \textbf{\bibinfo{volume}{17}},
  \bibinfo{pages}{033006} (\bibinfo{year}{2015}).
\bibitem[{\citenamefont{Astafiev et al}(2012)}]{ref14}
\bibinfo{author}{\bibfnamefont{O. V.}~\bibnamefont{Astafiev, et al.,}} \bibnamefont{}
\bibinfo{journal}{Nature} \textbf{\bibinfo{volume}{484}},
  \bibinfo{pages}{355-358} (\bibinfo{year}{2012}).
\bibitem[{\citenamefont{Arutyunov et al}(2012)}]{ref15}
\bibinfo{author}{\bibfnamefont{K. Y.}~\bibnamefont{Arutyunov, et al.,}} \bibnamefont{}
\bibinfo{journal}{Scientific Reports} \textbf{\bibinfo{volume}{2}},
  \bibinfo{pages}{293} (\bibinfo{year}{2012}).
\bibitem[{\citenamefont{Peltonen et al}(2013)}]{ref16}
\bibinfo{author}{\bibfnamefont{J. T.}~\bibnamefont{Peltonen, et al.,}} \bibnamefont{}
\bibinfo{journal}{Phys. Rev. B} \textbf{\bibinfo{volume}{88}},
  \bibinfo{pages}{220506} (\bibinfo{year}{2013}).
\bibitem[{\citenamefont{Peltonen et al}(2016)}]{ref17}
\bibinfo{author}{\bibfnamefont{J. T.}~\bibnamefont{Peltonen, et al.,}} \bibnamefont{}
\bibinfo{journal}{Phys. Rev. B} \textbf{\bibinfo{volume}{94}},
  \bibinfo{pages}{180508(R)} (\bibinfo{year}{2016}).
\bibitem[{\citenamefont{Svetogorov et~al.}(2018)\citenamefont{Svetogorov, Taguchi, Tokura,
  Basko, and Hekking}}]{ref18}
\bibinfo{author}{\bibfnamefont{A. E.}~\bibnamefont{Svetogorov}},
  \bibinfo{author}{\bibfnamefont{M.}~\bibnamefont{Taguchi}},
  \bibinfo{author}{\bibfnamefont{Y.}~\bibnamefont{Tokura}},
  \bibinfo{author}{\bibfnamefont{D.~M.} \bibnamefont{Basko}}, \bibnamefont{and}
  \bibinfo{author}{\bibfnamefont{F. W. J.}~\bibnamefont{Hekking}},
  \bibinfo{journal}{Phys. Rev. B} \textbf{\bibinfo{volume}{45}},
  \bibinfo{pages}{104514} (\bibinfo{year}{2018}).
\bibitem[{\citenamefont{Yoneda et~al.}(2011)\citenamefont{Yoneda, Niwa and Motohashi}}]{ref19}
\bibinfo{author}{\bibfnamefont{M.}~\bibnamefont{Yoneda}},
  \bibinfo{author}{\bibfnamefont{M.}~\bibnamefont{Niwa}}, \bibnamefont{and}
  \bibinfo{author}{\bibfnamefont{M.}~\bibnamefont{Motohashi}},
  \bibinfo{journal}{arxiv.org. cond-mat.mes-hall. 1108.3258.}
\bibitem[{\citenamefont{Yoneda et~al.}(2012)\citenamefont{Yoneda, Niwa and Motohashi}}]{ref20}
\bibinfo{author}{\bibfnamefont{M.}~\bibnamefont{Yoneda}},
  \bibinfo{author}{\bibfnamefont{M.}~\bibnamefont{Niwa}}, \bibnamefont{and}
  \bibinfo{author}{\bibfnamefont{M.}~\bibnamefont{Motohashi}},
  \bibinfo{journal}{Physica Scripta} \textbf{\bibinfo{volume}{2012}},
  \bibinfo{pages}{T151} (\bibinfo{year}{2012}).
\bibitem[{\citenamefont{Yoneda et~al.}(2014)\citenamefont{Yoneda, Obata and Niwa}}]{ref21}
\bibinfo{author}{\bibfnamefont{M.}~\bibnamefont{Yoneda}},
  \bibinfo{author}{\bibfnamefont{S.}~\bibnamefont{Obata}}, \bibnamefont{and}
  \bibinfo{author}{\bibfnamefont{M.}~\bibnamefont{Niwa}},
  \bibinfo{journal}{Mater. Trans} \textbf{\bibinfo{volume}{55}},
  \bibinfo{pages}{1510-1512} (\bibinfo{year}{2014}).
\bibitem[{\citenamefont{Yoneda et~al.}(2015)\citenamefont{Yoneda, Obata, Niwa and Motohashi}}]{ref22}
\bibinfo{author}{\bibfnamefont{M.}~\bibnamefont{Yoneda}},
  \bibinfo{author}{\bibfnamefont{S.}~\bibnamefont{Obata}},
  \bibinfo{author}{\bibfnamefont{M.}~\bibnamefont{Niwa}}, \bibnamefont{and}
  \bibinfo{author}{\bibfnamefont{M.}~\bibnamefont{Motohashi}},
  \bibinfo{journal}{Trans. Mat. Res. Soc. Japan} \textbf{\bibinfo{volume}{40[2]}},
  \bibinfo{pages}{115-118} (\bibinfo{year}{2015}).
\bibitem[{\citenamefont{Wees}(1991)}]{ref23}
\bibinfo{author}{\bibfnamefont{B. J.}~\bibnamefont{van Wees}},
  \bibinfo{journal}{Rev. B}
  \textbf{\bibinfo{volume}{44}},
 \bibinfo{pages}{28} (\bibinfo{year}{1991}).
\bibitem[{\citenamefont{Cha et~al.}(1991)\citenamefont{Cha, Fisher, Girvin, Wallin and Young}}]{ref24}
\bibinfo{author}{\bibfnamefont{M-C.}~\bibnamefont{Cha}},
  \bibinfo{author}{\bibfnamefont{M. P. A.}~\bibnamefont{Fisher}},
  \bibinfo{author}{\bibfnamefont{S. M.}~\bibnamefont{Girvin}},
  \bibinfo{author}{\bibfnamefont{M.} \bibnamefont{Wallin}}, \bibnamefont{and}
  \bibinfo{author}{\bibfnamefont{A. P.}~\bibnamefont{Young}},
  \bibinfo{journal}{Phys. Rev. B} \textbf{\bibinfo{volume}{44}},
  \bibinfo{pages}{6883} (\bibinfo{year}{1991}).
\bibitem[{\citenamefont{Wallin et~al.}(1994)\citenamefont{Wallin, S{\o}rensen, Girvin, and Young}}]{ref25}
\bibinfo{author}{\bibfnamefont{M.}~\bibnamefont{Wallin}},
\bibinfo{author}{\bibfnamefont{E. S.}~\bibnamefont{S{\o}rensen}},
  \bibinfo{author}{\bibfnamefont{S. M.}~\bibnamefont{Girvin}}, \bibnamefont{and}
  \bibinfo{author}{\bibfnamefont{A. P.}~\bibnamefont{Young}},
  \bibinfo{journal}{Phys. Rev. B} \textbf{\bibinfo{volume}{49}},
  \bibinfo{pages}{12115} (\bibinfo{year}{1994}).
\bibitem[{\citenamefont{Baturina and Vinokur}(2013)}]{ref26}
\bibinfo{author}{\bibfnamefont{T. I.}~\bibnamefont{Baturina}} \bibnamefont{and}
  \bibinfo{author}{\bibfnamefont{V. M.}~\bibnamefont{Vinokur}},
  \bibinfo{journal}{Ann. Phys} \textbf{\bibinfo{volume}{331}},
  \bibinfo{pages}{236-257} (\bibinfo{year}{2013}).
\bibitem[{\citenamefont{Diamantini et~al.}(2018)\citenamefont{Diamantini, Gammaitoni, Trugenberger and Vinokur}}]{ref27}
\bibinfo{author}{\bibfnamefont{M. C.}~\bibnamefont{Diamantini}},
\bibinfo{author}{\bibfnamefont{L.}~\bibnamefont{Gammaitoni}},
  \bibinfo{author}{\bibfnamefont{C. A.}~\bibnamefont{Trugenberger}}, \bibnamefont{and}
  \bibinfo{author}{\bibfnamefont{V. M.}~\bibnamefont{Vinokur}},
  \bibinfo{journal}{Scientific Reports} \textbf{\bibinfo{volume}{8}},
  \bibinfo{pages}{15718} (\bibinfo{year}{2018}).
\bibitem[{\citenamefont{Diamantini et~al.}(2018)\citenamefont{Diamantini, Gammaitoni, Trugenberger and Vinokur}}]{ref28}
\bibinfo{author}{\bibfnamefont{M. C.}~\bibnamefont{Diamantini}},
\bibinfo{author}{\bibfnamefont{L.}~\bibnamefont{Gammaitoni}},
  \bibinfo{author}{\bibfnamefont{C. A.}~\bibnamefont{Trugenberger}}, \bibnamefont{and}
  \bibinfo{author}{\bibfnamefont{V. M.}~\bibnamefont{Vinokur}},
 \bibinfo{journal}{Journal of Superconductivity and Novel Magnetism}\textbf{\bibinfo{volume}{32}},
  \bibinfo{pages}{47-51} (\bibinfo{year}{2019}).
\bibitem[{\citenamefont{Villain}(1975)}]{ref29}
\bibinfo{author}{\bibfnamefont{J.}~\bibnamefont{Villain}},
  \bibinfo{journal}{J. de Phys}
  \textbf{\bibinfo{volume}{36}},
 \bibinfo{pages}{581} (\bibinfo{year}{1975}).
\bibitem[{\citenamefont{Kleinert}(1989)}]{ref30}
\bibinfo{author}{\bibfnamefont{H.}~\bibnamefont{Kleinert}},
  \emph{\bibinfo{title}{Gauge fields in condensed matter. Vol. 1: Superflow and vortex lines. Disorder fields, phase transitions}}
  (\bibinfo{publisher}{World Scientific, Singapore}, \bibinfo{year}{1989}).
\bibitem[{\citenamefont{Janke and Kleinert}(1986)}]{ref31}
\bibinfo{author}{\bibfnamefont{W.}~\bibnamefont{Janke}} \bibnamefont{and}
  \bibinfo{author}{\bibfnamefont{H.}~\bibnamefont{Kleinert}},
  \bibinfo{journal}{Nuclear Physics B} \textbf{\bibinfo{volume}{270}},
  \bibinfo{pages}{135-153} (\bibinfo{year}{1986}).
\bibitem[{\citenamefont{Suzuki}(1976)}]{ref32}
\bibinfo{author}{\bibfnamefont{M.}~\bibnamefont{Suzuki}},
  \bibinfo{journal}{Commun.Math}
  \textbf{\bibinfo{volume}{51}},
 \bibinfo{pages}{183} (\bibinfo{year}{1976}).
\bibitem[{\citenamefont{Sondhi et~al.}(1997)\citenamefont{Sondhi, Girvin, Carini, and Shahar}}]{ref33}
\bibinfo{author}{\bibfnamefont{S. L.}~\bibnamefont{Sondhi}},
  \bibinfo{author}{\bibfnamefont{S. M.} \bibnamefont{Girvin}},
  \bibinfo{author}{\bibfnamefont{J. P.}~\bibnamefont{Carini}}, \bibnamefont{and}
  \bibinfo{author}{\bibfnamefont{D.}~\bibnamefont{Shahar}},
  \bibinfo{journal}{Rev. Mod. Phys} \textbf{\bibinfo{volume}{69}},
  \bibinfo{pages}{315} (\bibinfo{year}{1997}).
\bibitem[{\citenamefont{Hutchinson et~al.}(2015)\citenamefont{Hutchinson, Keating and Mezzadri}}]{ref34}
\bibinfo{author}{\bibfnamefont{J.}~\bibnamefont{Hutchinson}},
  \bibinfo{author}{\bibfnamefont{J. P.}~\bibnamefont{Keating}}, \bibnamefont{and}
  \bibinfo{author}{\bibfnamefont{F.}~\bibnamefont{Mezzadri}},
  \bibinfo{journal}{Advances in Mathematical Physics} \textbf{\bibinfo{volume}{2015}},
  \bibinfo{pages}{652026} (\bibinfo{year}{2015}).
\bibitem[{\citenamefont{Fazio et~al.}(1991)\citenamefont{Fazio and Schöni}}]{ref35}
\bibinfo{author}{\bibfnamefont{R.}~\bibnamefont{Fazio}}, \bibnamefont{and}
  \bibinfo{author}{\bibfnamefont{G.}~\bibnamefont{Schön}},
  \bibinfo{journal}{Phys. Rev. B} \textbf{\bibinfo{volume}{43}},
  \bibinfo{pages}{5307–5320} (\bibinfo{year}{1991}).
\bibitem[{\citenamefont{Diamantini, et~al.}(1996)\citenamefont{Diamantini, Sodano, and Trugenberger}}]{ref36}
\bibinfo{author}{\bibfnamefont{M. C.}~\bibnamefont{Diamantini}},
  \bibinfo{author}{\bibfnamefont{P.}~\bibnamefont{Sodano}}, \bibnamefont{and}
  \bibinfo{author}{\bibfnamefont{C. A.}~\bibnamefont{Trugenberger}},
  \bibinfo{journal}{Nucl. Phys. B} \textbf{\bibinfo{volume}{474}},
  \bibinfo{pages}{641-677} (\bibinfo{year}{1996}).
\bibitem[{\citenamefont{Diamantini, et~al.}(2018)\citenamefont{Diamantini, Trugenberger and Vinokur}}]{ref37}
\bibinfo{author}{\bibfnamefont{M. C.}~\bibnamefont{Diamantini}},
  \bibinfo{author}{\bibfnamefont{C. A.}~\bibnamefont{Trugenberger}}, \bibnamefont{and}
  \bibinfo{author}{\bibfnamefont{V. M.}~\bibnamefont{Vinokur}},
  \bibinfo{journal}{Nature Comm. Phys} \textbf{\bibinfo{volume}{1}},
  \bibinfo{pages}{Article number:77} (\bibinfo{year}{2018}).
\end{thebibliography}

\end{document}